\documentclass[12pt, letterpaper]{JHEP3}
\usepackage{latexsym,amssymb,amsmath,epsfig,epic,eepic}
\newcommand{\tr}{\mathop{\mathrm{tr}}\nolimits}

\newcommand{\beq}{\begin{equation}}
\newcommand{\eeq}{\end{equation}}
\newcommand{\beqs}{\begin{eqnarray}}
\newcommand{\eeqs}{\end{eqnarray}}

\newcommand{\nn}{\nonumber}

\newcommand{\Ncal}{\mathcal{N}}

\newcommand{\ov}{\overline}

\title{Instantons and  Toric Quiver Gauge Theories}

\author{\parbox{12cm}{Riccardo Argurio$^1$, Gabriele Ferretti$^2$
and  Christoffer Petersson$^2$}
\\
~\\
~\\
$^1$Physique Th\'eorique et Math\'ematique and
International Solvay Institutes \\ Universit\'e Libre de
Bruxelles,
CP 231, 1050 Bruxelles, Belgium\\

\vspace{0.3cm}
$^2$Department of Fundamental Physics\\ Chalmers
University of Technology, 412 96 G\"oteborg, Sweden
\vspace{0.3cm}

\email{rargurio@ulb.ac.be,  ferretti@chalmers.se, chrpet@chalmers.se}\\}

\abstract{We show how to construct the general action coupling
  (multi)instantons to gauge theories arising from branes probing arbitrary
  toric singularities. We give a general set of rules for how to
  construct such an action given the knowledge of the superpotential for the
  gauge theory. The main idea is to obtain the action by higgsing a theory
  whose instanton dynamics is known, namely an orbifold of $\mathcal{N}=4$
  super Yang-Mills. We find that the couplings of the fermionic zero-modes
  with the holomorphic fields are dictated by the structure of the
  superpotential describing the toric singularity. We present explicit
  examples such as the Suspended Pinch Point, the Conifold and the first three
  del Pezzo's. We perform various checks on these results by further higgsing
  to smaller orbifolds and present some applications, including both gauge
  theory and stringy instantons.}
\begin{document}
\setcounter{section}{0}
\renewcommand{\thefootnote}{\arabic{footnote}}
\setcounter{footnote}{0} \setcounter{page}{1}

\tableofcontents

\section{Introduction}
Instantons represent a class of non-perturbative phenomena in gauge theory and
string theory that is particularly amenable to theoretical study. In their
original incarnation~\cite{Belavin:1975fg} they were described by solutions to
the euclidean equations of motion of a gauge theory and their semiclassical analysis~\cite{'t Hooft:1976fv} exposed the existence of chiral symmetry violating terms in the effective action of great relevance to particle phenomenology.

After the advent of D-branes~\cite{Polchinski:1995mt}, it was soon realized that D-branes with a euclidean world-sheet wrapped on a non trivial cycle could give rise to closely related
effects~\cite{Witten:1995gx,Douglas:1995bn,Witten:1996bn,Ganor:1996pe,Green:2000ke,Billo:2002hm}.
(In perturbative string theory, fundamental strings with a euclidean world-sheet wrapped on such a cycle, known as world-sheet instantons, had already been extensively used~\cite{Dine:1986zy}.)

The analogy between D-brane instantons and ordinary gauge instantons is striking but there are also subtle differences. While a gauge instanton requires a gauge theory whose fields provide the necessary background, a D-brane instanton arises from a geometrical object that exists (in this framework) independently on the D-branes giving rise to the gauge theory, and it has colloquially speaking, a life of its own. This fact opens up the possibility of considering configurations that do not correspond to ordinary gauge theory instantons but nevertheless modify the gauge dynamics by giving rise to new terms in the effective lagrangian.
These configurations and their contributions are known as ``exotic" or
``stringy''. Their
properties have been intensely investigated in the last few years due to their
relevance to $\mathcal{N}=1$ dynamics, such as MSSM/GUT phenomenology, moduli
stabilization, and dynamical supersymmetry (SUSY) breaking 
\cite{Blumenhagen:2006xt}--\cite{Blumenhagen:2008ji}.

For a single instanton, one can roughly distinguish three types of configurations of interest:

If a euclidean D-brane wraps a cycle on which more than one space-filling D-branes are also wrapped, we are in a situation similar to that of an ordinary gauge instanton. In this case, one expects the generation of the familiar instanton induced corrections to the
superpotential~\cite{Taylor:1982bp,Affleck:1983mk} provided that the rank assignment of the various groups is the correct one.

If the euclidean D-brane wraps a cycle on which no space-filling D-brane is
present one is faced with an exotic configuration, without direct gauge theory
analogue. The study of this configuration is made
difficult~\cite{Argurio:2007vqa} by the presence of extra neutral fermionic
zero-modes for the instanton stemming from the fact that the instanton
spectrum is not sensitive to the presence of the space-filling D-brane and
thus it mantains all four fermionic goldstino zero-modes arising from breaking
half of the eight supercharges of a typical type II Calabi-Yau
compactification. In order to get a non-vanishing contribution to the
holomorphic quantities of the theory one is required to lift two such
fermionic modes and the most readily available tool to accomplish this task is
an orientifold projection \cite{Argurio:2007vqa,Franco:2007ii}.

The third case, where there is only one space-filling D-brane wrapping the same cycle as the instanton, is somewhat in the middle and is also very interesting. Although from the gauge theory point of view one does not expect any instanton solution in a $U(1)$ theory it can be shown that the presence of one space-filling brane is enough to soak-up the extra zero-modes and in some cases one gets contributions to the superpotential, once again provided the other rank assignments are correct~\cite{Petersson:2007sc} (see also~\cite{Aganagic:2007py,GarciaEtxebarria:2007zv}).

These three basic cases can of course be combined into more complex, multi-instanton configurations that display quite a rich structure. In this case various instantons can also split and recombine along curves of marginal stability~\cite{GarciaEtxebarria:2007zv}.

The study of D-brane instantons has taken place in different contexts, most notably the brane-world scenarios where one first compactifies space-time to four dimensions, then engineers a phenomenologically interesting $\mathcal{N}=1$ gauge theory with a configuration of wrapping and intersecting branes and orientifolds and finally generates non-perturbative effects by wrapping euclidean D-branes on the geometric cycles. But there is also great interest in considering ``local" constructions of D-branes probing a space-time singularity in an otherwise non-compact six dimensional manifold. This is of course crucial in the context of the gauge/gravity correspondence but it is also relevant to string phenomenology when properly embedded in a consistent configuration.

In order to make progress one has to have control over the action describing
the coupling of the fields in the gauge theory to the instanton moduli, as
well as the action describing the interaction of the moduli among
themselves. Since, for the time being, we are only interested in corrections
to the holomorphic quantities of the gauge theory (most notably the
superpotential), we will restrict ourselves to the coupling of the moduli
to chiral superfields. Their couplings can be derived by a variety of means,
the most direct one being applicable to the case of D-branes probing an
orbifold or orientifold singularity~\cite{Douglas:1996sw}, where a conformal field theory (CFT)
description is straightforwardly available.

It is however important to try to go beyond the orbifold limit, particularly
having in mind applications to the gauge/gravity correspondence, where
orbifold gauge theories provide too restrictive a class of models. In this
context it is much more interesting to consider gauge theories arising from
D-branes probing a toric singularity
\cite{Morrison:1998cs}--\cite{Bertolini:2004xf}, where the techniques
developed in the last decade provide a beautiful set of phenomena such as
Seiberg duality, cascades and, in some cases, confinement or dynamical SUSY
breaking \cite{Berenstein:2005xa}--\cite{Argurio:2006ny}. Theories arising from toric singularities also have the trademark of possessing chiral operators that, in spite of being non-renormalizable by naive power-counting, become exactly marginal in the infrared and contribute to the superpotential, in accordance to the AdS/CFT correspondence. We will see that these operators also play an important role in the instanton dynamics.

In this paper we address the above issue and show how to construct the general action coupling instantons to gauge theories arising from branes probing arbitrary toric singularities. We will give a general set of rules for how to construct such an action given the knowledge of the superpotential for the gauge theory. We will consider many explicit examples such as the Suspended Pinch Point (SPP), the Conifold and the first three del Pezzo's ($dP_1$, $dP_2$ and $dP_3$).

The basic idea behind our construction is the well known fact
(see e.g.~\cite{Morrison:1998cs}--\cite{Feng:2001xr}) that any quiver
gauge theory describing D-branes at a toric singularity can be obtained by
higgsing a sufficiently large orbifold, for which techniques are readily
available to obtain the instanton action. The higgsing procedure can be
applied (with some care) to the instanton sector as well yielding all the
desired couplings. This method works quite generally and it applies to rigid
instantons as well as instantons with internal neutral modes. In fact, the
role played by these extra neutral modes in the multi-instanton case is
crucial for the higgsing procedure to work. Indeed, as it will become clear,
single instantons in a toric geometry will generically descend
from multi-instantons in the unhiggsed parent theory. 

Although the main focus of
this paper is to present the general technique and some basic examples, we
will also briefly touch upon a few applications, such as the ubiquitous nature
of the Affleck-Dine-Seiberg (ADS) superpotential and various comments on
exotic contributions to the theory on the SPP and the del Pezzo's.

The paper is organized as follows:

In section~2 we outline the general strategy
of our approach and spell out the rules that can be used to obtain the
instanton action for a gauge theory arising from D-branes at a toric singularity.
We do this in a way that will hopefully allow the reader, who is not interested in going through the lengthy algebraic arguments, to construct the coupling needed in the specific case of interest. The remainder of the paper is essentially a justification of these rules with examples and applications.

Section~3 is a short summary of the well known properties of the $\mathcal{N}=4$ theory and its orbifolds needed in our construction.

Section~4 is the first and simplest example of how the higgsing procedure works for the instanton. Although nothing new is learned in this case, since one goes from a well known model (the $\mathcal{N}=2~$ $\mathbb{C}^2/\mathbb{Z}_2$ orbifold) to another well known case (the $\mathcal{N}=4$ theory itself), this shows in detail how we will apply the procedure to more complicated cases and should also be thought of as a first consistency check.

Section~5 discusses the construction of the instanton actions for SPP and the Conifold from the higgsing of the $\mathbb{C}^3/\mathbb{Z}_2 \times \mathbb{Z}_2$ orbifold, together with two further brief consistency checks.
Already at this stage one sees the deviation of the instanton action from the
one for an orbifold gauge theory. Namely there exist higher order holomorphic couplings of the charged fermionic zero-modes to the chiral superfields that follow the same index pattern as the superpotential. These terms are required for consistency with further higgsing and cannot be neglected.

In section~6 we take a short break and discuss the recovery of the ADS superpotential from the charged bosonic and the fermionic anti-holomorphic couplings. This result is well known but put in this context it shows the necessity of not altering the anti-holomorphic couplings with respect to the naive expectations from the orbifold theory. This fact is consistent with our findings.

Section~7 continues discussing more examples, namely the first three del Pezzo's as embedded in a $\mathbb{C}^3/\mathbb{Z}_3 \times \mathbb{Z}_3$ orbifold. These models have some intrinsic interest in the context of dynamical SUSY breaking but we also discuss them at length because, contrary to all models discussed in section~5, they are chiral and we would like to show that our procedure works in this case as well. We hope the reader will not be put off by the long formulas in this section and will remember that they \emph{also} follow straightforwardly from the rules of section~2.

In section~8 we end by giving some sample computations of stringy instanton effects.
We will be rather sketchy, since we hope to return to these applications in a separate publication.

While this work was in progress, we became aware of the related but complementary work appearing in \cite{Kachru:2008wt}.

\section{General strategy}
In this section we outline the general strategy
of our approach and spell out the rules that can be used to obtain the
instanton action for a gauge theory arising from D-branes at a toric singularity.

We start by recalling that it is always possible to embed the toric diagram
describing such a singularity into that of a sufficiently large orbifold
singularity. From the gauged
linear sigma model (GLSM) description of the singularity, it is possible to see
that this means one can go from the orbifold singularity to the non-orbifold
one by partial resolutions, i.e. by turning on some Fayet-Iliopoulos (FI) terms in the GLSM.
From the quiver gauge theory point of view, turning on background FI
terms\footnote{As is usual in the literature, we will assume that all
the statements which are strictly correct when all nodes of the quiver
are $U(1)$ classical gauge groups carry over smoothly to the general case
where the nodes are (strongly coupled) $SU(N)$s.} necessitates that
some fields acquire vacuum expectation values (VEVs) in order to satisfy the D-flatness conditions.
As a consequence, by the Brout-Englert-Higgs mechanism, some pairs of
gauge groups will be higgsed to their diagonal part, eating in the process
the field which had a VEV. Some other matter fields then acquire a mass
and are integrated out, typically yielding new terms in the superpotential
which are of order higher than cubic (of course, orbifolds only have
cubic superpotentials).

We propose to apply this procedure also to determine the structure of instanton
zero modes for D-branes on non-orbifold toric singularities. Namely, we would
like to apply the higgsing procedure not only to the quiver gauge theory,
but to the quiver gauge theory coupled to its instanton sector.

Recall that quite generically the zero modes of a (multi)-instanton configuration can be divided into
neutral (associated to open strings stretched between two Euclidean D-branes) and charged
(associated to open strings with one end on the instanton and one on the gauge theory brane).
The charged zero modes are those that couple directly to the fields in the gauge theory.
It is clear that the matter field VEVs will give masses to some of the charged instanton zero
modes. Moreover, a corresponding neutral zero mode
must also obtain a VEV and correspondingly some neutral zero modes will also become massive. This can be understood in several ways. Firstly, it is clear that there cannot be more kinds of instantons than gauge groups
in a given geometry, since the number of gauge groups is essentially
given by the number of non trivial compact cycles over which D-branes
can wrap, before any anomaly argument is put forward. Secondly,
the same background closed string mode which
generates the FI term in the matter sector also generates a similar
term in the instanton neutral zero mode sector, which is (before the ADHM
limit \cite{Billo:2002hm}) just the reduction to zero dimensions of the quiver gauge theory.
As a consequence, the structure of the surviving neutral zero modes
mirrors exactly the structure of quiver gauge and matter fields. This means that if we give a VEV to the chiral superfield $\Phi_{ab}$ in the bi-fundamental representation of the gauge groups associated to node $a$ and $b$ we will also give a VEV to the corresponding neutral scalar zero mode $s_{ab}$, a complex combination of the real zero modes, usually denoted by $\chi$, present in the non rigid case.

Taking all the VEVs into account, one goes on to see
which charged instanton zero modes become massive and how they couple to the chiral fields of the gauge theory
and to the neutral zero modes. One finds that only the ``diagonal" zero modes (connecting a gauge group of the quiver with its own instanton) and the fermionic zero modes with an index structure similar to the chiral superfields surviving
in the quiver remain.

More explicitly:
\begin{itemize}
\item Consider a specific node $a$ of a particular quiver gauge theory. There
  will always be charged bosonic  and fermionic zero modes connecting this
  node with its own instanton node, also denoted by $a$. According to the
  traditional notation such zero modes will be denoted by
  $\omega_{\dot\alpha,aa}$, (a bosonic ``spinor" - the index $\dot \alpha=1,2$
  will very often be omitted in writing), $\mu_{aa}$, (a fermionic ``scalar") and by their conjugates $\bar\omega_{\dot\alpha,aa}$ and $\bar\mu_{aa}$ going from the instanton node to the gauge theory node.
\item  Consider now two specific nodes $a$ and $b$ (not necessarily distinct)
  of a particular quiver gauge theory. For each chiral superfield $\Phi_{ab}$
  connecting these two nodes there will be a corresponding fermionic zero mode
  $\mu_{ab}$ connecting the gauge group $a$ to the instanton $b$ and its
  ``conjugate" $\bar\mu_{ab}$, this time connecting the instanton $a$ to the gauge group $b$. Notice that the two zero modes are described by two completely distinct arrows in the extended quiver which may be chiral.
\end{itemize}
As an illustration of these rule consider the trivial case of $\mathcal{N}=4$. The quiver consists of a single node with three incoming/outgoing arrows corresponding to the three chiral superfields of the $\mathcal{N}=1$ notation. There will be thus one set of bosonic modes $(\omega_{\dot\alpha}, \bar\omega_{\dot\alpha})$, and four sets of fermionic modes
$(\mu^A, \bar\mu^A)$, ($A=1,2,3,4$), one from the first rule and three from the second one.

So far this is a straightforward generalization of the rules for the orbifold theory and it is as expected.
However some care is needed in integrating out
the massive zero modes to obtain the couplings. For the generic quiver gauge theory, instead of generalizing the CFT techniques used in the case of orbifolds, we rely on other methods. The two main tools we use are the splitting of the fermionic instanton action into a holomorphic and a anti-holomorphic piece and the consistency condition that, if by further higgsing we recover a smaller orbifold, the action obtained by these rules must match the well known one obtained by CFT.

The results we obtain are quite simple to express and seem to be completely generic.
\begin{itemize}
\item The coupling between the bosonic charged zero modes
and the superfields, as well as the anti-holomorphic coupling between the charged
fermionic zero modes and the superfields is exactly
as in the orbifold case.
Namely, for every pair of nodes $a$ and $b$ for which the relevant fields exist there will be the following couplings:
\beq
      \bar\omega_{aa}\Phi_{ab}\Phi^\dagger_{ba}\omega_{aa}, \quad
      \bar\omega_{aa}\Phi^\dagger_{ab}\Phi_{ba}\omega_{aa}, \quad
      \bar\mu_{aa}\Phi^\dagger_{ab}\mu_{ba}, \quad
      \bar\mu_{ab}\Phi^\dagger_{ba}\mu_{aa}.
\eeq
In the case of multiple instantons there are similar couplings between the charged moduli and the neutral moduli, denoted by $s$, that are crucial for consistency and will be discussed at length in the following. However the extra neutral moduli will not be present in the case of a single (fractional) instanton and this is the configuration that is mostly studied in practical applications.
\item  The holomorphic coupling between the charged fermionic
zero modes and the superfields is
obtained by taking each term in the superpotential and, while keeping
the same quiver index structure, substituting two fermionic charged zero
modes and all combinations of matter fields $\Phi$ and neutral bosonic zero modes $s$ allowed by the symmetries.
Again ignoring the bosonic modes $s$ for the time being, this rule means that, if one encounters, say, the term $\tr \Phi_{12} \Phi_{23} \Phi_{34} \Phi_{41}$ in the superpotential, one must expect the four terms:
\beq
   \tr \big( \bar\mu_{12} \Phi_{23} \Phi_{34} \mu_{41} + \bar\mu_{23} \Phi_{34} \Phi_{41} \mu_{12} +
   \bar\mu_{34} \Phi_{41} \Phi_{12} \mu_{23} +   \bar\mu_{41} \Phi_{12} \Phi_{23} \mu_{34} \big)
\eeq
in the instanton action.
\end{itemize}

In the following we will also see that the bosonic neutral modes can be accommodated in the same way, remembering to put them in a different position (to the \emph{right} of $\mu$) due to their different index structure. Also, as it will become clear, there is a relative factor of $(-1)$ for each $s$ appearing in the \emph{holomorphic} part.\footnote{As a result of higgsing in the instanton sector, there will also be
couplings higher than cubic among neutral zero modes only. We will
not focus on them in the following, since they follow straightforwardly
from the reduction to zero dimensions of the action of the toric quiver.
These couplings can play a crucial role in some multi-instanton
configurations, see \cite{GarciaEtxebarria:2007zv}. Notice however that some of them will be
eventually suppressed in the ADHM limit.}

The remainder of the paper is a justification and test of the above rules. The reader who is not interested in the algebraic details and is willing to take these rules for granted can simply skim through the notation in the next section and look at the few examples and applications in sections~6 and~8.

\section{Notation and conventions}

In this short chapter we review the notation for the well known orbifold case
that will serve as a starting point in our analysis.
We start from a gauge theory living on D-branes probing a simple orbifold of $\mathbb{C}^3$. From the
perturbative, open string point of view, the quiver gauge theory is
just obtained in the following way. One formulates ${\cal N}=4$
SYM in ${\cal N}=1$ language and assigns to each of its fields
a Chan-Paton structure derived from the orbifold projection. Specifically, since we will only consider abelian orbifolds, the structure of the gauge superfields turns out to be block diagonal, each block denoting the node of a quiver. The chiral superfields $\Phi^i$ will have some components set to zero by the orbifold projection and the remaining submatrices $\Phi_{ab}$ will transform in the bi-fundamental (or adjoint if $a$=$b$) representation. This
already determines the quiver. The remaining data is encoded in the
superpotential following directly from inserting these matrices of fields
into the cubic ${\cal N}=4$ SYM superpotential~\footnote{We have written the 6 scalars $X_a$ of ${\cal N}=4$ SYM
first as an $SU(4)$ antisymmetric matrix $X_{AB}=(\Sigma^a)_{AB}X_a$
and then we have identified $\Phi^i=\frac{1}{2}\epsilon^{ijk} X_{jk}$
and $\Phi^\dagger_i=X_{i4}$.}
\beq
W_{{\cal N}=4} = \tr \Phi^1 [ \Phi^2 , \Phi^3], \label{wN4}
\eeq

In the instanton sector, one can again start from the
spectrum and couplings of the instanton zero modes for ${\cal N}=4$ SYM,
which is well-known and can also be computed straightforwardly in
perturbation theory.

In order to set the stage for the rest of the paper
and get acquainted with the different kinds of zero modes, we write
here the action of the zero modes:
\begin{eqnarray}
S_1 & = & \tr\Big\{-[a_\mu,s^{\dagger}_{i}] [a^\mu,s^i ]-
\frac{i}{2} \Big( M^{\alpha i} [s^{\dagger}_{i}, M^4_\alpha]- \frac{1}{2}\epsilon_{ijk}M^{\alpha i} [s^{j}, M^k_\alpha] \Big) \nn \\
 &  & \phantom{\tr } + i \left(\bar\mu^i \omega_{\dot\alpha} +
\bar\omega_{\dot\alpha} \mu^i + \sigma^\mu_{\beta
\dot\alpha}{[M^{\beta i}, a_\mu]}\right)\! \lambda^{\dot\alpha}_i
 + i \left(\bar\mu^4 \omega_{\dot\alpha} +
\bar\omega_{\dot\alpha} \mu^4 + \sigma^\mu_{\beta
\dot\alpha}{[M^{\beta 4}, a_\mu]}\right)\! \lambda^{\dot\alpha}_4 \nn \\
&&\phantom{\tr } - i D^c\!\left( \bar\omega^{\dot \alpha}
(\tau^c)^{\dot\beta}_{\dot\alpha} \omega_{\dot\beta} + i
\bar\eta^c_{\mu\nu}  {[a^\mu, a^\nu]}\right) \!\Big\}, \label{s1}
\end{eqnarray}
In the expression (\ref{s1}) $a_\mu$, $s^i$ and $D^a$ are neutral bosonic zero modes,
$M^A_\alpha$ and $\lambda^{\dot \alpha}_A$ (with $A=i,4$) are fermionic
neutral zero modes, while $\omega_{\dot \alpha}$ and
$\bar \omega_{\dot \alpha}$ are bosonic charged zero modes and
$\mu^A$, $\bar \mu^A$ are charged fermionic zero modes~\footnote{Recall that
neutral modes are those corresponding to strings with both ends on an
instanton, while the charged ones are those with one end on an instanton and
the other on a spacetime filling D-brane, i.e. they are in the
(anti)fundamental of a gauge group.}. The three complex fields $s^i$ are the complexification of the six real zero modes usually denoted by $\chi$ and they prove more convenient for the formulations of the interactions.

To the above action, we must add the terms that couple the charged zero
modes to the matter fields. We find it convenient to write it
together with terms that we omitted in (\ref{s1}), which
couple the charged zero modes to the bosonic neutral zero modes:
\begin{eqnarray}
\label{s2}
S_2 & = & \tr\Big\{ \frac{1}{2}
\Big( \bar\omega_{\dot\alpha }\Phi^i+s^i \bar\omega_{\dot\alpha }  \Big) \Big(\Phi^\dagger_i \omega^{\dot\alpha }+ \omega^{\dot\alpha } s^{\dagger}_{i}  \Big) + \frac{1}{2}\Big(  \bar\omega_{\dot\alpha }  \Phi^\dagger_i+s^{\dagger}_{i}\bar\omega_{\dot\alpha }  \Big)\Big( \Phi^i \omega^{\dot\alpha }+\omega^{\dot\alpha }s^i  \Big) \nn \\
&&\phantom{\tr\Big\{ }+\frac{i}{2} \bar\mu^i \big(\Phi^\dagger_i \mu^4 + \mu^4 s^{\dagger}_{i}\big)
- \frac{i}{2} \bar\mu^4 \big(  \Phi^\dagger_i \mu^i +\mu^i s^{\dagger}_{i} \big)
-\frac{i}{2} \epsilon_{ijk} \bar\mu^i \big( \Phi^j \mu^k - \mu^j s^k \big)  \Big\}~. \nn \\
\end{eqnarray}
In order to be complete, we must also write the terms in the action
that are actually suppressed in the ADHM limit \cite{Billo:2002hm},
but which here play
a role in lifting some of the neutral modes:
\beqs
S_3 & = & \tr\Big\{ \frac{1}{2} D_c^2 -
\frac{i}{2} \Big( \lambda_{\dot \alpha i} [s^{i},
\lambda^{\dot \alpha}_{4}]- \frac{1}{2}\epsilon^{ijk}\lambda_{\dot \alpha i}
[s^\dagger_{j}, \lambda^{\dot \alpha}_{k}] \Big) \nonumber \\
& & \phantom{\tr\Big\{ }
+ [ s^i, s^j] [ s^\dagger_j,s^\dagger_i] + \frac{1}{2}
[ s^i,s^\dagger_i][s^j ,  s^\dagger_j] \label{sadhm}
\Big\}.
\eeqs

As for the gauge theory, taking the orbifold projection the instanton zero modes too will become larger
matrices of zero modes carrying Chan-Paton indices relating them
to the different instantons and gauge groups. The rules determining which components survive the projection are straightforward and we shall follow the conventions discussed in \cite{Argurio:2007vqa}. Namely, $s^i, M^i, \lambda_i, \mu^i, \bar\mu^i$ acquire the same structure as $\Phi^i$, whereas $a_\mu, D_c, \omega, \bar\omega, M^4, \lambda_4, \mu^4, \bar\mu^4$ are all block diagonal.

The action for the zero modes is again found by substitution in the ${\cal N}=4$ zero mode actions (\ref{s1}), (\ref{s2}), (\ref{sadhm}). After reducing the supersymmetry we have the important option of adding a FI term which modifies the very last term of (\ref{sadhm}) by letting:
\beq
\label{FIinclude}
  [ s^i,s^\dagger_i]_{aa} \to [ s^i,s^\dagger_i]_{aa} - \xi_{aa}
  \equiv \sum_{\langle b a \rangle} (s^\dagger_{ab} s_{ba} - s_{ab} s^\dagger_{ba}) - \xi_{aa},
\eeq
where $\xi$ is a block diagonal matrix and the sum is over the nodes $b$ connected to $a$ by a line in the quiver.

\section{Warm-up: Higgsing from  $\mathbb{C}^2/\mathbb{Z}_2$ to ${\cal N}=4$}
We start by performing a higgsing procedure in both the matter
and the instanton moduli sectors in a set up where we know from perturbative
string theory both the starting point and the end point. Namely we will go from the simplest of all orbifolds $\mathbb{C}^2/\mathbb{Z}_2$, yielding a ${\cal N}=2$ theory down to the ${\cal N}=4$ theory by resolving the singularity and higgsing one chiral field. There will be no suprises, but this exercise is useful to adjust the whole procedure so that it yields consistent results.

\begin{figure}
\begin{center}
\includegraphics[width=110mm]{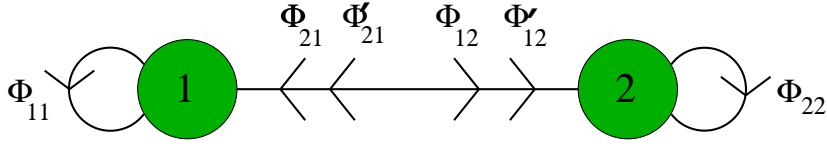}
\caption{\small{The quiver diagram for the $\mathbb{Z}_2$ theory.}}
\label{Z2D}
\end{center}
\end{figure}

We thus start by spelling out the field content of the
$\mathbb{C}^2/\mathbb{Z}_2$ quiver gauge theory, including the instanton
sector. It is simply obtained recalling that the orbifold action
acts as $g: (z_1,z_2,z_3) \rightarrow (z_1,-z_2,-z_3)$ and that
its representation on the Chan-Paton indices is given by $\gamma(g)\equiv
\sigma_3$. As reviewed in the previous section we get a block-diagonal gauge field (each block
can be as usual considered a $N_i\times N_j$ matrix in general) and
the matter fields are
\beq
\Phi^1 =\begin{pmatrix}
   \Phi_{11}   &    \\
      &  \Phi_{22}
      \end{pmatrix},\quad
\Phi^2 = \begin{pmatrix}
      & \Phi_{12}    \\
  \Phi_{21}     &
\end{pmatrix},\quad
\Phi^3 = \begin{pmatrix}
      &  \Phi^\prime_{12}   \\
   \Phi^\prime_{21}    &
\end{pmatrix}~, \label{phiz2}
\eeq
see the quiver diagram in Figure \ref{Z2D}.

The superpotential is given by (Note, from here on we will omit writing the trace explicitly):
\beq
   W_{\mathbb{Z}_2} =\Phi^\prime_{21} \Phi_{11} \Phi_{12} - \Phi_{21} \Phi_{11} \Phi^\prime_{12} +
                       \Phi^\prime_{12} \Phi_{22} \Phi_{21} - \Phi_{12}
                       \Phi_{22} \Phi^\prime_{21}    \label{wz2}
\eeq
just by replacing (\ref{phiz2}) in the ${\cal N}=4$ expression (\ref{wN4}).

In the instanton sector, we first consider the bosonic neutral modes,
which are generally matrices constituted by $k_i\times k_j$ matrix blocks,
where $k_i$ can be considered as the instanton number in the gauge group
of each node.
Then the $a_\mu$ are block diagonal as the $A_\mu$ gauge fields, and the
$s^i$ have the same form as the $\Phi^i$.

The projection of the fermionic neutral modes has also been reviewed in the previous section and turns out to be
\beqs &&
M^1 =\begin{pmatrix}
   M_{11}   &    \\
      &  M_{22}
      \end{pmatrix},\quad
M^2 = \begin{pmatrix}
      & M_{12}    \\
  M_{21}     &
\end{pmatrix},\nonumber \\ &&
M^3 = \begin{pmatrix}
      &  M^\prime_{12}   \\
   M^\prime_{21}    &
\end{pmatrix},\quad
M^4 =\begin{pmatrix}
   M^\prime_{11}   &    \\
      &  M^\prime_{22}
      \end{pmatrix} \label{fermZ2}
\eeqs
where we have suppressed the spinor index $\alpha$. The structure of
the $\lambda^{\dot\alpha}_A$ zero modes is exactly the same as above.

\begin{figure}
\begin{center}
\includegraphics[width=130mm]{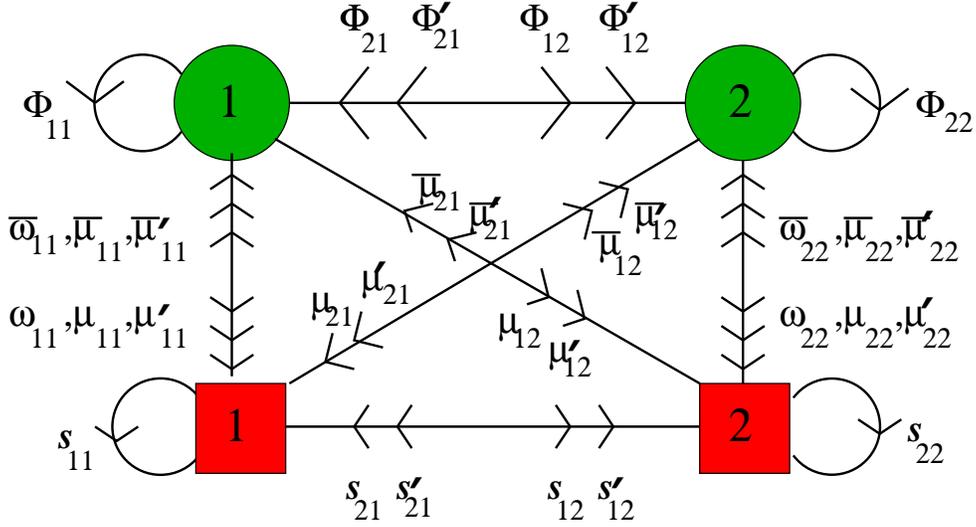}
\caption{\small{The $\mathbb{Z}_2$ theory with instantons.}}
\label{Z2wInstD}
\end{center}
\end{figure}

For the charged instanton zero modes, we can decompose the
matrices into $k_i\times N_j$ and $N_i\times k_j$ blocks. The bosonic
modes $\omega_{\dot\alpha}$ and $\bar \omega_{\dot\alpha}$ are
block diagonal, while the fermionic modes $\mu^A$ and $\bar \mu^A$
have the same form as in (\ref{fermZ2}) above. From now on we will
denote all the modes by indices relating them to the relevant
instanton and/or gauge nodes, see Figure \ref{Z2wInstD}.

The complete action for the zero modes above, and their coupling to the
matter fields, is simply given by plugging
back the above definitions into the ${\cal N}=4$ action given in the
previous section.

We can now perform the higgsing, which corresponds to resolving
the $\mathbb{C}\times (\mathbb{C}^2/\mathbb{Z}_2)$ singularity to (locally)
$\mathbb{C}^3$. In the quiver, this is achieved by giving a VEV
proportional to the identity to a bifundamental field:
\beq
\Phi'_{21}=m. \label{vevz2}
\eeq
Of course, this requires the two gauge groups to have the same rank
$N_1=N_2\equiv N$. Moreover, giving a VEV to a single field is consistent
only if we turn on (opposite) FI terms for the diagonal $U(1)$ factors
of both nodes, $\xi_2=-\xi_1=|m|^2$.

\begin{figure}
\begin{center}
\includegraphics[width=25mm]{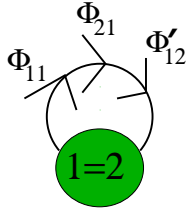}
\caption{\small{The $\mathcal{N}=4$ theory higgsed down from the $\mathbb{Z}_2$ theory.}}
\label{N4fZ2D}
\end{center}
\end{figure}

From the superpotential (\ref{wz2}), we see that such a VEV gives a
mass to the fields $\Phi_{12}$ and $\Phi_{11}-\Phi_{22}$, which can
then be integrated out. The F-term for $\Phi_{12}$ sets $\Phi_{11}=\Phi_{22}$
exactly, so that we are left with the 3 matter fields $\Phi_{11}$,
$\Phi_{21}$, $\Phi'_{12}$ and 2 terms in the superpotential which reproduce the $\mathcal{N}=4$ superpotential
(\ref{wN4}), see Figure \ref{N4fZ2D}.

\begin{figure}
\begin{center}
\includegraphics[width=40mm]{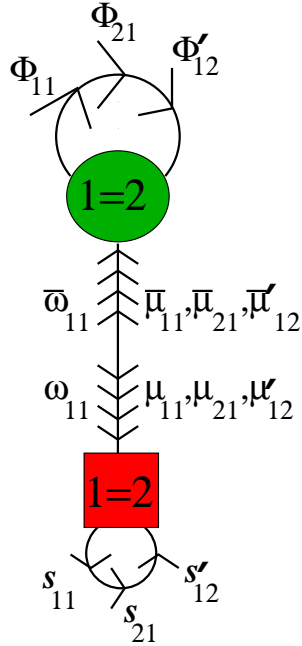}
\caption{\small{The ${\cal N}=4$ theory with instantons higgsed from the $\mathbb{Z}_2$ theory.}}
\label{N=4fZ2wInstD}
\end{center}
\end{figure}

We can now start to consider the effect of the VEV (\ref{vevz2}) on the
instanton zero modes, see Figure \ref{N=4fZ2wInstD}.
 First of all, we consider the coupling of the matter
fields to the bosonic charged zero modes $\omega_{\dot\alpha}$ and
$\bar \omega_{\dot\alpha}$. The relevant piece of the instanton action, if we do not include the neutral $s$ fields, reads:
\beq
S_2 \supset  \frac{1}{2} \Big( \bar \omega_{11} \Phi'^{\dagger}_{12}
\Phi'_{21} \omega_{11} +  \bar \omega_{22} \Phi'_{21}\Phi'^{\dagger}_{12}
\omega_{22} \Big)= \frac{1}{2}|m|^2  (\bar \omega_{11} \omega_{11} +
\bar \omega_{22} \omega_{22})~,
\eeq
where we have suppressed the $\dot \alpha$ indices and denoted
$(\Phi_{ij})^\dagger=\Phi^\dagger_{ji}$. We see that, as it stands above, a single VEV would actually give a mass and
lift all the charged bosonic zero modes. This is clearly not what we expect,
since of course there should be one pair of charged bosonic zero modes
in the ${\cal N}=4$ theory we obtain after higgsing.

In order to recover this result, we see that we also have to consider the
coupling to the neutral bosonic zero modes $s'_{21}$:
\beq
S_2 \supset  \frac{1}{2} \Big(  \bar \omega_{11} \Phi'^{\dagger}_{12}
+s'^\dagger_{12} \bar \omega_{22}\Big) \Big(\Phi'_{21} \omega_{11}
+\omega_{22} s'_{21}\Big)
+ \frac{1}{2} \Big( \bar \omega_{22} \Phi'_{21}
+s'_{21} \bar \omega_{11}\Big) \Big(\Phi'^{\dagger}_{12}\omega_{22}
+\omega_{11} s'^\dagger_{12}\Big).
\eeq
It is clear that if we give a VEV
\beq
s'_{21}=-m \label{svev}
\eeq
then the above action becomes a mass term for only one linear combination
of the zero modes
\beq
S_2 \supset  |m|^2 (  \bar \omega_{11} -  \bar \omega_{22})
( \omega_{11} - \omega_{22}).
\eeq
That the neutral bosonic zero mode acquires a VEV such as (\ref{svev})
can be understood as follows. Recall that in order to give a VEV
to the matter field, we have to turn on FI terms. Those are actually associated
to turning on a background value for a closed string (twisted) modulus.
As the disk amplitudes with spacefilling or euclidean boundaries are
very much alike, we expect that a FI term
will also appear in the D-term-like piece of the action for the $s_i$
as given in (\ref{FIinclude}). Consequently, the action will be mininized
by giving a VEV to the bosonic zero mode in (\ref{svev}).
Note also that the F-term like piece of (\ref{sadhm}) will in turn produce
mass terms that lift the zero modes $s_{12}$ and $s_{11}-s_{22}$,
exactly in the same way as it happens in the matter sector.\footnote{
We see that in order for the latter zero modes to be consistently lifted,
the ADHM scaling limit that suppresses (\ref{sadhm}) has to be performed
after the VEV, or FI parameter, is turned on.
In other words, the VEV will have to
eventually scale in the ADHM limit in such a way that the masses for
the zero modes that we have integrated out do not vanish.}

Taking now $\bar \omega_{11} =  \bar \omega_{22}$, $\omega_{11} =
\omega_{22}$, $s_{11}=s_{22}$ and $s_{12}=0$ (the latter equality
can be seen as a consequence of the scaling limit, see the discussion later
on), we see that the action coupling the bosonic zero modes and the matter
fields is exactly the one for the ${\cal N}=4$  theory, as in the
first line of (\ref{s2}), after we make the identifications
$\Phi_{11}\equiv \Phi_{22}=\frac{1}{\sqrt{2}} \Phi^1$, $\Phi_{21}=\Phi^2$,
$\Phi'_{12}=\Phi^3$ and similarly for the $s^i$. (The factors of $\sqrt{2}$ are necessary in order to keep all the fields canonically normalized.)

We now turn to consider the fermionic charged zero modes $\bar \mu$ and $\mu$.
Inserting the VEVs (\ref{vevz2}) and (\ref{svev}) in (\ref{s2}), we obtain the following
``mass'' terms: (dropping an overall normalization factor)
\beq
S_2 \supset m^* \bar\mu'_{21}(\mu'_{11}-\mu'_{22}) + m^*
(\bar\mu'_{11}-\bar\mu'_{22})\mu'_{21}
+m \bar\mu_{12}(\mu_{11}-\mu_{22}) + m
(\bar\mu_{11}-\bar\mu_{22})\mu_{12}~.
\eeq

In order to integrate out these 8 zero modes, one should perform a
Gaussian integral. This involves a non-trivial determinant,
since there are other terms in (\ref{s2}), involving the above fermionic zero modes coupling to matter fields and neutral bosonic
zero modes. However, one
easily realizes that in order to match the result of this integration
with the ${\cal N}=4$ result, one has to set to zero (i.e. scale away)
all terms that have a prefactor which is at least $1/|m|^2$. This is
what we will indeed do here and in the following, but note that we are
nevertheless allowing for the possibility of keeping terms which go
as some power of $1/m$ or of $1/m^*$, i.e. are holomorphic or anti-holomorphic
in $m$. The reason why we want to keep them will be clear in the next
section, where we perform two or more consecutive higgsings.
Why ``holomorphic'' terms do not scale away while non-holomorphic ones do
cannot be rigorously justified in the present set up, but is presumably
related to the fact that the former are protected while the
latter receive large corrections during the non-trivial RG flow that
the theory undergoes from its classical description discussed here and
its IR effective dynamics.

As we are interested only in the corrections proportional to
$1/m$ or $1/m^*$, we can integrate out independently the two sets
of modes, by setting in turn $1/m=0$ and $1/m^*=0$. Doing this, it turns
out that everything works in the present case
as if we could set all the ``massive'' zero modes
to zero, hence imposing $\mu_{11}=\mu_{22}$ and
$\mu'_{11}=\mu'_{22}$ exactly, and similarly for the barred ones.

Performing the identifications as before, together with
$\mu_{11}=\frac{1}{\sqrt{2}}\mu^1$, $\mu_{21}=\mu^2$, $\mu'_{12}=\mu^3$,
$\mu'_{11}=\frac{1}{\sqrt{2}}\mu^4$ and similarly for $\bar\mu^A$,
we obtain exactly the couplings in the second line of (\ref{s2}),
up to a global prefactor of $\frac{1}{\sqrt{2}}$ which can be
reabsorbed by performing a further overall rescaling.

We are left to discuss the bosonic neutral zero modes $a_\mu$ and
the fermionic ones $M^A$ and $\lambda_A$. All the relevant components of
these zero modes are lifted by the VEV of the zero mode $s'_{21}$
exactly in the same way as the VEV for $\Phi'_{21}$ lifts the gauge
fields and the gaugini superpartners of the fields which become massive.
For instance, the components $M_{11}-M_{22}$, $M'_{11}-M'_{22}$, $M'_{21}$
and $M_{12}$ will be lifted through the couplings in the last two terms
in the first line of (\ref{s1}).
Similarly, the components $\lambda_{11}-\lambda_{22}$,
$\lambda'_{11}-\lambda'_{22}$, $\lambda'_{12}$ and $\lambda_{21}$
will get a mass
through the couplings in the last two terms
in the first line of (\ref{sadhm}). Note that this too implies that
the ADHM limit has to be taken in such a way that these mass terms
are not washed away. Eventually, the first term in the first line
of (\ref{s1}) gives a mass to the combination $a_{\mu 11}-a_{\mu 22}$,
leaving the center-of-mass bosonic zero modes as the ones relevant
for the instantons in the $\Ncal =4$ theory.\footnote{The auxiliary
terms $D^a$ are already massive. The ones which are related to
$a_\mu$ zero modes that have become massive can be integrated out
trivially by setting them to zero.}

As a last routine check, one can reexpress the last two lines of (\ref{s1})
in terms of the zero modes that have been kept and recover the
$\Ncal =4$ expression.

We have thus addressed in this section all the subtleties related
to the higgsing procedure in the instanton sector which are already
present when one is going from one orbifold singularity to another.
In the following section we can thus address the additional features
that appear when one exits the realm of orbifold singularities.

\section{Higgsing $\mathbb{C}^3/\mathbb{Z}_2\times\mathbb{Z}_2$ to the
Suspended Pinch Point and further}
We now address the first instance of higgsing to a non-orbifold toric
geometry, where it is less direct to compute the spectrum and action
by perturbative methods.

There is a major difference between orbifold and non-orbifold
(toric) quivers, in the sense that orbifold quivers are conformal
at the classical and perturbative level, while the non-orbifold quivers
are typically non-conformal classically (there are terms higher than cubic
in the superpotential) but possess a non-trivial superconformal fixed point
at finite coupling. Hence non orbifold quivers are defined by the classical
field content and superpotential up to this RG flow to the IR fixed point.
We conjecture here that the structure of the instanton moduli does not
change along this flow. That this is a consistent thing to do is checked
by higgsing back to some other orbifold quiver, and recovering the
spectrum and couplings computed in perturbation theory.

We will start from the $\mathbb{C}^3/\mathbb{Z}_2\times\mathbb{Z}_2$
four node quiver, and higgs it to the Suspended Pinch Point (SPP) three
node quiver. Then we will further higgs the latter to the previously
discussed $\mathbb{C}^2/\mathbb{Z}_2$ two node quiver, as a consistency
check. We will also higgs the SPP to the  Conifold
quiver, to gain confidence in the structure of the moduli action
in the non-orbifold case. Consistency is again checked by further higgsing the Conifold
 to recover the ${\cal N}=4$ theory.

Note that the only non generic feature of the quiver gauge theories
discussed in this section is that they are non-chiral. We will
address the more general chiral theories later on, but we anticipate
that there will be no additional features as far as the construction of the instanton action is concerned.

We begin by concisely reviewing the structure of the
$\mathbb{C}^3/\mathbb{Z}_2\times\mathbb{Z}_2$ quiver gauge theory
and of its instanton zero modes, see~\cite{Morrison:1998cs,Bertolini:2001gg}. (This orbifold has been 
recently investigated from the point of view of string phenomenology in~\cite{Berg:2005ja}). As in \cite{Argurio:2007vqa}, the orbifold
is taken as $g_1: (z_1,z_2,z_3) \rightarrow (z_1,-z_2,-z_3)$ and
$g_2: (z_1,z_2,z_3) \rightarrow (-z_1,z_2,-z_3)$, with the
representation on the Chan-Paton indices being
$\gamma(g_1)=\sigma_3 \otimes \mathbf{1}$ and
$\gamma(g_2)=\mathbf{1} \otimes\sigma_3 $.
Then, the gauge fields and the 3 matter fields are
given by
\beqs
A^\mu = \begin{pmatrix} A^\mu_{11} & 0 &
0   & 0   \cr  0 & A^\mu_{22} & 0   & 0 \cr 0 & 0   & A^\mu_{33}   &  0 \cr  0
& 0   & 0 & A^\mu_{44} \cr\end{pmatrix}, &&
\Phi^1 = \begin{pmatrix}0 & \Phi_{12} &
0   & 0   \cr  \Phi_{21} & 0 & 0   & 0 \cr 0 & 0   & 0   &  \Phi_{34} \cr  0
& 0   & \Phi_{43} & 0 \cr\end{pmatrix},
\nonumber \\
\Phi^2 = \begin{pmatrix}0 & 0
& \Phi_{13} &  0  \cr     0   & 0   & 0   & \Phi_{24} \cr \Phi_{31} & 0   & 0
& 0  \cr  0   & \Phi_{42} & 0 & 0 \cr\end{pmatrix}, && \Phi^3 =
\begin{pmatrix}0 & 0   & 0   & \Phi_{14} \cr  0   & 0 & \Phi_{23} & 0   \cr
0 & \Phi_{32} & 0   & 0   \cr  \Phi_{41} & 0   & 0   & 0 \cr\end{pmatrix}~.
\label{structure}
\eeqs

As already discussed in full generality, in the instanton sector
$s^i, M^i, \lambda_i, \mu^i, \bar\mu^i$ acquire the same structure as
$\Phi^i$, whereas $a_\mu, D_c, \omega, \bar\omega, M^4, \lambda_4,
\mu^4, \bar\mu^4$ are all block diagonal as $A_\mu$.

\begin{figure}
\begin{center}
\includegraphics[width=65mm]{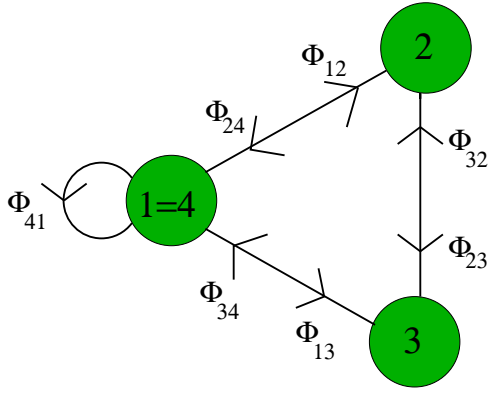}
\caption{\small{The SPP theory higgsed down from the $\mathbb{Z}_2\times\mathbb{Z}_2$ theory.}}
\label{SPPD}
\end{center}
\end{figure}

The superpotential for the $\mathbb{C}^3/\mathbb{Z}_2\times\mathbb{Z}_2$
quiver is simply
\beqs  W_{\mathbb{Z}_2 \times \mathbb{Z}_2} &=&
\Phi_{31}\Phi_{12}\Phi_{23}-
\Phi_{31}\Phi_{14}\Phi_{43}-\Phi_{13}\Phi_{32}\Phi_{21}+\Phi_{13}\Phi_{34}\Phi_{41} \nonumber \\ &&
+\Phi_{42}\Phi_{21}\Phi_{14}-
\Phi_{42}\Phi_{23}\Phi_{34}-\Phi_{24}\Phi_{41}\Phi_{12}+\Phi_{24}\Phi_{43}\Phi_{32}~.
         \label{wZ2Z2}
\eeqs
We can now see what happens if we give a VEV such as
\beq
\Phi_{14} = m.
\eeq
This requires the condition $N_1 =N_4 =N$ and breaks the two gauge groups corresponding to nodes 1 and 4 to the diagonal subgroup,
\begin{equation}
SU(N)_1 \times SU(N)_4 \to SU(N)_{(14)}~.
\end{equation}
The chiral superfield $\Phi_{41}$ will thereby transform in the adjoint representation of $SU(N)_{(14)}$.
We immediately see that the fields $\Phi_{31}$, $\Phi_{43}$, $\Phi_{42}$
and $\Phi_{21}$ become massive. One should integrate them out through
their F-flatness equations, which read:
\beqs
\Phi_{31}=\frac{1}{m}\Phi_{32}\Phi_{24}, &&
\Phi_{43}=\frac{1}{m}\Phi_{12}\Phi_{23}, \nn \\
\Phi_{42}=\frac{1}{m}\Phi_{13}\Phi_{32}, &&
\Phi_{21}=\frac{1}{m}\Phi_{23}\Phi_{34} ~.\label{intout}
\eeqs
Inserting these values back into (\ref{wZ2Z2}) gives us the SPP superpotential,
\beq  W_{SPP} =
\frac{1}{m}\Phi_{24}\Phi_{12}\Phi_{23}\Phi_{32}-\frac{1}{m}\Phi_{13}\Phi_{32}\Phi_{23}\Phi_{34}
+\Phi_{13}\Phi_{34}\Phi_{41}-\Phi_{24}\Phi_{41}\Phi_{12}
         \label{wSPP}
\eeq
where all the remaining fields (except for $\Phi_{41}$) transform in bifundamental representations of two of the factors in the gauge group $SU(N)_{(14)}\times SU(N_2) \times SU(N_3 )$, see Figure \ref{SPPD}.

\begin{figure}
\begin{center}
\includegraphics[width=100mm]{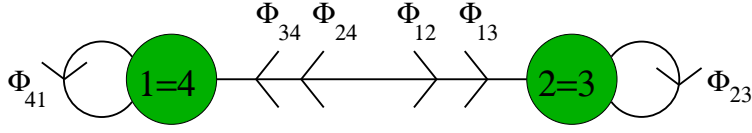}
\caption{\small{The $\mathbb{Z}_2$ theory higgsed down from the SPP theory.}}
\label{Z2hD}
\end{center}
\end{figure}

We can continue with this procedure and obtain the quiver gauge theory for a $\mathbb{Z}_2$ orbifold if we start from the SPP theory and give an additional VEV to the chiral superfield $\Phi_{32}$,
\begin{equation}
\Phi_{32} = m ~.
\label{vev32}
\end{equation}
This means that we have the condition $N_3 =N_2 =M$ and that we ``pinch'' the two gauge groups corresponding to nodes 2 and 3 together,
\begin{equation}
SU(M)_2 \times SU(M)_3 \to SU(M)_{(23)}~.
\end{equation}
As before, the chiral superfield $\Phi_{23}$ will now transform in the adjoint representation of SU$(M)_{(23)}$. We see from (\ref{wSPP}) that (\ref{vev32}) does not induce any new mass terms, but gives us the $\mathbb{Z}_2$ superpotential, (same as (\ref{wz2}) upon relabeling)
\beq
W_{\mathbb{Z}_2} =
\Phi_{24}\Phi_{12}\Phi_{23}-\Phi_{13}\Phi_{23}\Phi_{34}
+\Phi_{13}\Phi_{34}\Phi_{41}-\Phi_{24}\Phi_{41}\Phi_{12}
\eeq
where $\Phi_{12}$, $\Phi_{13}$ are in the ($\square$,$\overline{\square}$) of the gauge group $SU(N)_{(14)}\times SU(M)_{(23)}$, while $\Phi_{24}$, $\Phi_{34}$ are in the ($\overline{\square}$,$\square$) and $\Phi_{41}$, $\Phi_{23}$ are in the adjoint of the respective gauge groups, see Figure \ref{Z2hD}.

\begin{figure}
\begin{center}
\includegraphics[width=70mm]{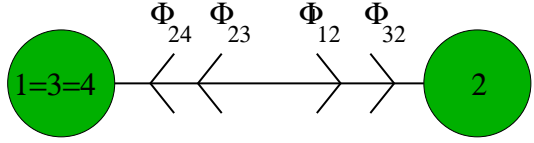}
\caption{\small{The Conifold theory higgsed down from the SPP theory.}}
\label{ConiD}
\end{center}
\end{figure}

To get the conifold gauge theory, we start again from the SPP theory but now we instead give a VEV to the chiral superfield $\Phi_{34}$,
\begin{equation}
\Phi_{34} = m .
\label{vev34}
\end{equation}
This implies the condition $N_1 =N_4 =N_3 =N$, such that the three gauge groups corresponding to nodes 1, 3 and 4 now become
\begin{equation}
SU(N)_1 \times SU(N)_3 \times SU(N)_4 \to SU(N)_{(134)}~.
\end{equation}
 We see from (\ref{wSPP}) that (\ref{vev34}) induces a mass term for the bifundamental chiral superfield $\Phi_{13}$ and the adjoint field $\Phi_{41}$. Hence, we solve for these fields and get the following expressions,
\begin{equation}
\Phi_{13}=\frac{1}{m}\Phi_{12}\Phi_{24}, \qquad
\label{int34}
\Phi_{41}=\frac{1}{m}\Phi_{32}\Phi_{23}~.
\end{equation}
Inserting (\ref{vev34}) and (\ref{int34}) into (\ref{wSPP}) yields the superpotential for the conifold,
\beqs  W_{con} &=&
\frac{1}{m}\Phi_{12}\Phi_{23}\Phi_{32}\Phi_{24}-\frac{1}{m}\Phi_{12}\Phi_{24}\Phi_{32}\Phi_{23}
\label{wConifold}
\eeqs
where $\Phi_{12}$, $\Phi_{32}$ are in the ($\square$,$\overline{\square}$) of
the gauge group $SU(N)_{(134)}\times SU(N_2)$, $\Phi_{24}$, $\Phi_{23}$ are in
the ($\overline{\square}$,$\square$) and there are no more adjoint fields, see Figure \ref{ConiD}.

\begin{figure}
\begin{center}
\includegraphics[width=25mm]{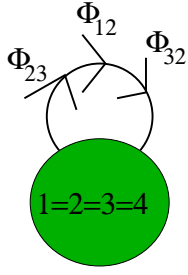}
\caption{\small{The $\mathcal{N}=4$ theory higgsed down from the Conifold theory.}}
\label{N4fZ2Z2D}
\end{center}
\end{figure}

From here, giving a VEV to, say, $\Phi_{24}=m$, leads straightforwardly
to the $\Ncal=4$ theory and its cubic superpotential, see Figure \ref{N4fZ2Z2D}.

All the above is of course standard, but we will now see how this
higgsing pattern extends to the instanton sector.

Let us consider our first non-trivial step out of the orbifold realm.
As reviewed in the previous section, in the absence of any other field
acquiring a VEV, the effect that $\Phi_{14}=m$ would produce is to give
a mass to both pairs of bosonic charged moduli $\omega_{11}$,
$\bar\omega_{11}$ and $\omega_{44}$, $\bar \omega_{44}$. We know that
in order to lift only one combination of these two pairs of zero modes,
we need to turn on the VEV
\beq
s_{14}=-m.
\eeq
We will then have
\beqs
S_2 &\supset &  \frac{1}{2} \Big(  \bar \omega_{11} \Phi_{14}
+s_{14} \bar \omega_{44}\Big) \Big(\Phi^\dagger_{41} \omega_{11}
+\omega_{44} s^\dagger_{41}\Big)
+ \frac{1}{2} \Big( \bar \omega_{44} \Phi^\dagger_{41}
+s^\dagger_{41} \bar \omega_{11}\Big) \Big(\Phi_{14}\omega_{44}
+\omega_{11} s_{14}\Big)\nn\\
& = &
 |m|^2 (  \bar \omega_{11} -  \bar \omega_{44})
( \omega_{11} - \omega_{44}).
\eeqs
We thus see that we can set $\bar \omega_{11} =  \bar \omega_{44}$
and $\omega_{11} = \omega_{44}$. The action coupling the bosonic
charged zero modes to the matter fields and to the neutral bosonic
zero modes is obtained as follows. We see that if we replace
the fields that we have integrated out (\ref{intout}) by their values
in the action for the orbifold zero modes, we would get a series of terms
which are quartic in the matter fields and have a $\frac{1}{|m|^2}$
prefactor. Similarly, we know that because of the F-terms in (\ref{sadhm})
the zero modes $s_{31}$, $s_{43}$, $s_{42}$ and $s_{21}$ will
be integrated out, with expressions such as $s_{31}=-\frac{1}{m}s_{32}s_{24}$.
Hence there would also be terms quartic in the $s^i$, and terms
such as $\bar \omega \Phi\Phi \omega s^\dagger s^\dagger$. However, all of
these terms have the same $\frac{1}{|m|^2}$ prefactor, and we will
assume, based on consistency with further higgsing, that these
terms are suppressed due to the RG flow that essentially decouples
the massive modes (both in the matter and in the instanton sectors).

Hence, we can write the following action:
\beqs
S^{\omega}_{SPP}&=&   \big( \bar \omega_{11} \Phi_{12}
+s_{12} \bar \omega_{22}\big) \big(\Phi^\dagger_{21} \omega_{11}
+\omega_{22} s^\dagger_{21}\big)
+ \big( \bar \omega_{22} \Phi^\dagger_{21}
+s^\dagger_{21} \bar \omega_{11}\big) \big(\Phi_{12}\omega_{22}
+\omega_{11} s_{12}\big)\nonumber
\\ &
  +& \big( \bar \omega_{11} \Phi_{13}
+s_{13} \bar \omega_{33}\big) \big(\Phi^\dagger_{31} \omega_{11}
+\omega_{33} s^\dagger_{31}\big)
+ \big( \bar \omega_{33} \Phi^\dagger_{31}
+s^\dagger_{31} \bar \omega_{11}\big) \big(\Phi_{13}\omega_{33}
+\omega_{11} s_{13}\big)\nonumber\\ &
  +& \big( \bar \omega_{22} \Phi_{24}
+s_{24} \bar \omega_{11}\big) \big(\Phi^\dagger_{42} \omega_{22}
+\omega_{11} s^\dagger_{42}\big)
+ \big( \bar \omega_{11} \Phi^\dagger_{42}
+s^\dagger_{42} \bar \omega_{22}\big) \big(\Phi_{24}\omega_{11}
+\omega_{22} s_{24}\big)\nonumber\\ &
  +& \big( \bar \omega_{33} \Phi_{34}
+s_{34} \bar \omega_{11}\big) \big(\Phi^\dagger_{43} \omega_{33}
+\omega_{11} s^\dagger_{43}\big)
+ \big( \bar \omega_{11} \Phi^\dagger_{43}
+s^\dagger_{43} \bar \omega_{33}\big) \big(\Phi_{34}\omega_{11}
+\omega_{33} s_{34}\big)\nonumber\\ &
  +& \big( \bar \omega_{22} \Phi_{23}
+s_{23} \bar \omega_{33}\big) \big(\Phi^\dagger_{32} \omega_{22}
+\omega_{33} s^\dagger_{32}\big)
+ \big( \bar \omega_{33} \Phi^\dagger_{32}
+s^\dagger_{32} \bar \omega_{22}\big) \big(\Phi_{23}\omega_{33}
+\omega_{22} s_{23}\big)\nonumber \\ &
  +&\big( \bar \omega_{33} \Phi_{32}
+s_{32} \bar \omega_{22}\big) \big(\Phi^\dagger_{23} \omega_{33}
+\omega_{22} s^\dagger_{23}\big)
+ \big( \bar \omega_{22} \Phi^\dagger_{23}
+s^\dagger_{23} \bar \omega_{33}\big) \big(\Phi_{32}\omega_{22}
+\omega_{33} s_{32}\big)\nonumber \\ &
  +&\big( \bar \omega_{11} \Phi_{41}
+s_{41} \bar \omega_{11}\big) \big(\Phi^\dagger_{14} \omega_{11}
+\omega_{11} s^\dagger_{14}\big)
+ \big( \bar \omega_{11} \Phi^\dagger_{14}
+s^\dagger_{14} \bar \omega_{11}\big) \big(\Phi_{41}\omega_{11}
+\omega_{11} s_{41}\big) \nn \\
\label{SboseSPP}
\eeqs

In presenting the above actions and in all the following ones we find ourself facing a notational dilemma.
The higgsing procedure removes some of the nodes making some of the indices obsolete. One could relabel the indices at every step, for instance in the case under consideration letting $4 \to 1$ everywhere. The advantage of doing this is that the instanton action for each model looks more intelligible but the disadvantage is that this relabeling makes it difficult to follow the chain of higgsings from one model to another. We choose \emph{not} to relabel the fields at this stage and ask the reader to keep track of which nodes are identified. We will however still set $\omega_{44}=\omega_{11}$ etc. because these are two previously distinct moduli fields that are now identified by the higgsing procedure.

As we turn to consider the charged fermionic zero modes, we see
that the VEVs for $\Phi_{14}$ and $s_{14}$ yield the following
mass terms:
\beq
S_2 \supset  m^* \big[ \bar \mu_{14} (\mu_{11}-\mu_{44})
+ (\bar\mu_{11}-\bar\mu_{44})\mu_{14} \big]
 + m \big(\bar \mu_{21} \mu_{42} + \bar \mu_{42} \mu_{21}
- \bar \mu_{31} \mu_{43} - \bar \mu_{43} \mu_{31}\big)
\label{massmu}
\eeq
As in the previous section, the complete action coupling $\mu, \bar \mu$
to $\Phi$ and $s$ will contain other terms  involving the above
massive zero modes. These would lead to very complicated expressions mixing the holomorphic/anti-holomorphic components. Again, consistency is obtained if and only if we decouple
all terms with a $\frac{1}{|m|^2}$ prefactor, thus preserving the holomorphic/anti-holomorphic serparation. The remaining terms
are computed by setting in turn $\frac{1}{m}=0$ and $\frac{1}{m^*}=0$ and reinstating the finite value of $m$ in the end, after the integration.

If we set $\frac{1}{m}=0$, and thus set exactly to zero all the modes
which have a mass $m$, we observe that  no  couplings remain which
involve $ \mu_{14}$ or $\bar \mu_{14}$. Hence, $\mu_{11}-\mu_{44}$
and $\bar\mu_{11}-\bar\mu_{44} $ act like Lagrange multipliers and
integrating out the modes with a mass $m^*$ is equivalent to setting
to zero all terms in which they appear.

On the other hand setting $\frac{1}{m^*}=0$, we set exactly to zero the
modes with mass $m^*$, but this still leaves terms which mix the
remaining massless modes
with the modes of mass $m$. Hence integrating out the latter
yields non trivial expressions. Yet it is remarkable that these expressions
only contain {\em holomorphic} dependence on the fields $\Phi$ and
$s$.

It is easy to see that because of the structure of the terms involving
$\Phi^\dagger$ and $s^\dagger$, and due to the previous remarks,
no terms linear in $\frac{1}{m^*}$ will be generated at all.
Hence, the part of the action coupling the fermionic charged zero modes
to the anti-holomorphic fields will remain cubic:
\beqs
S_{SPP}^{\overline{\mathrm{holo}}} &=& \big(\bar \mu_{41} \Phi^\dagger_{14}
 +\bar \mu_{12} \Phi^\dagger_{21} + \bar \mu_{13} \Phi^\dagger_{31}\big)
 \mu_{11} + \big(\bar \mu_{23} \Phi^\dagger_{32}
 +\bar \mu_{24} \Phi^\dagger_{42}\big) \mu_{22}  \nn \\
&& +  \big(\bar \mu_{32} \Phi^\dagger_{23}
 +\bar \mu_{34} \Phi^\dagger_{43}\big) \mu_{33}
+  \big( s^\dagger_{14} \bar \mu_{41} +  s^\dagger_{42} \bar \mu_{24} +
 s^\dagger_{43} \bar \mu_{34} \big) \mu_{11} \nn \\
&&+  \big( s^\dagger_{21} \bar \mu_{12} +
s^\dagger_{23} \bar \mu_{32} \big) \mu_{22}
+  \big( s^\dagger_{31} \bar \mu_{13} +
s^\dagger_{32} \bar \mu_{23} \big) \mu_{33} \nn \\ &&
-\bar\mu_{11} \big( \Phi^\dagger_{14} \mu_{41} + \Phi^\dagger_{42} \mu_{24} +
\Phi^\dagger_{43} \mu_{34} \big)
 -\bar\mu_{22} \big( \Phi^\dagger_{21} \mu_{12}
+ \Phi^\dagger_{23} \mu_{32} \big)\nn \\ &&
 -\bar\mu_{33} \big( \Phi^\dagger_{31} \mu_{13}
+ \Phi^\dagger_{32} \mu_{23} \big)
-\bar\mu_{11} \big( \mu_{41} s^\dagger_{14} + \mu_{12} s^\dagger_{21} +
\mu_{13} s^\dagger_{31} \big) \nn \\ &&
-\bar\mu_{22} \big( \mu_{24} s^\dagger_{42} + \mu_{23} s^\dagger_{32} \big)
-\bar\mu_{33} \big( \mu_{34} s^\dagger_{43} + \mu_{32} s^\dagger_{23} \big) \label{SantiSPP}
\eeqs
The expressions for the charged fermionic zero modes of mass $m$ which
have been integrated out are
\beqs
\mu_{31}= \frac{1}{m}\big( \Phi_{32} \mu_{24} - \mu_{32} s_{24}\big), &\quad&
\bar \mu_{31}= \frac{1}{m}\big( \bar\mu_{32}\Phi_{24} - s_{32}\bar\mu_{24}\big),
\nn \\
\mu_{43}= \frac{1}{m}\big( \Phi_{12} \mu_{23} - \mu_{12} s_{23}\big), &\quad&
\bar \mu_{43}= \frac{1}{m}\big( \bar\mu_{12}\Phi_{23} - s_{12}\bar\mu_{23}\big),
\nn \\
\mu_{42}= \frac{1}{m}\big( \Phi_{13} \mu_{32} - \mu_{13} s_{32}\big), &\quad&
\bar \mu_{42}= \frac{1}{m}\big( \bar\mu_{13}\Phi_{32} - s_{13}\bar\mu_{32}\big),
\nn \\
\mu_{21}= \frac{1}{m}\big( \Phi_{23} \mu_{34} - \mu_{23} s_{34}\big), &\quad&
\bar \mu_{21}= \frac{1}{m}\big( \bar\mu_{23}\Phi_{34} - s_{23}\bar\mu_{34}\big).
\eeqs
One can actually replace all the terms where the above massive modes appear
by substituting the above expressions in the mass terms of the second line
of (\ref{massmu}).
This will clearly lead to terms like $\bar\mu\Phi\Phi\mu$, $\bar\mu\mu s s$
and $\bar \mu \Phi \mu s$. Other terms like $\bar\mu\Phi\Phi\mu$ and
 $\bar\mu\mu s s$ are generated when replacing the expressions such as
 (\ref{intout}) for the
$\Phi$ and $s$ modes that are also integrated out. All in all,
we arrive at the following expression for the holomorphic couplings
of the fermionic charged zero modes:
\beqs
    S_{SPP}^{\mathrm{holo}} &=& \frac{1}{m}(\bar\mu_{24}\mu_{12}s_{23}s_{32} - \bar\mu_{24}\Phi_{12}\mu_{23}s_{32} +
    \bar\mu_{24}\Phi_{12}\Phi_{23}\mu_{32}+ \bar\mu_{12}\mu_{23}s_{32}s_{24} \nonumber \\&& - \bar\mu_{12}\Phi_{23}\mu_{32}s_{24} +
    \bar\mu_{12}\Phi_{23}\Phi_{32}\mu_{24}+ \bar\mu_{23}\mu_{32}s_{24}s_{12} - \bar\mu_{23}\Phi_{32}\mu_{24}s_{12}  \nonumber \\&&+
    \bar\mu_{23}\Phi_{32}\Phi_{24}\mu_{12} + \bar\mu_{32}\mu_{24}s_{12}s_{23} - \bar\mu_{32}\Phi_{24}\mu_{12}s_{23} +
    \bar\mu_{32}\Phi_{24}\Phi_{12}\mu_{23})  \nonumber \\&&- \frac{1}{m}(\bar\mu_{13}\mu_{32}s_{23}s_{34} - \bar\mu_{13}\Phi_{32}\mu_{23}s_{34} +
    \bar\mu_{13}\Phi_{32}\Phi_{23}\mu_{34} + \bar\mu_{32}\mu_{23}s_{34}s_{13}  \nonumber \\&&- \bar\mu_{32}\Phi_{23}\mu_{34}s_{13} +
    \bar\mu_{32}\Phi_{23}\Phi_{34}\mu_{13} + \bar\mu_{23}\mu_{34}s_{13}s_{32} - \bar\mu_{23}\Phi_{34}\mu_{13}s_{32}  \nonumber \\&&+
    \bar\mu_{23}\Phi_{34}\Phi_{13}\mu_{32} + \bar\mu_{34}\mu_{13}s_{32}s_{23} - \bar\mu_{34}\Phi_{13}\mu_{32}s_{23} +
    \bar\mu_{34}\Phi_{13}\Phi_{32}\mu_{23})  \nonumber \\&&- \bar\mu_{13}\mu_{34}s_{41} + \bar\mu_{13}\Phi_{34}\mu_{41} -\bar\mu_{34}\mu_{41}s_{13} + \bar\mu_{34}\Phi_{41}\mu_{13} \nonumber \\&& - \bar\mu_{41}\mu_{13}s_{34} + \bar\mu_{41}\Phi_{13}\mu_{34} + \bar\mu_{24}\mu_{41}s_{12} - \bar\mu_{24}\Phi_{41}\mu_{12}
     \nonumber \\&&+ \bar\mu_{41}\mu_{12}s_{24} - \bar\mu_{41}\Phi_{12}\mu_{24}  + \bar\mu_{12}\mu_{24}s_{41} - \bar\mu_{12}\Phi_{24}\mu_{41} \label{SholoSPP}
\eeqs

The rules described in section~2 should be clear by comparing the above action with the expression for the superpotential (\ref{wSPP}). A term of order four in the superpotential gives rise to twelve terms in the holomorphic instanton action, obtained by inserting $\bar\mu$ and $\mu$ in $4\times 3$ ways and closing the trace with $\Phi$ or $s$ accordingly. Similarly, each cubic term gives rise to $3 \times 2$ terms.

At this point we are ready to make a consistency check. If the procedure we followed is correct, by further higgsing $\Phi_{32} = - s_{32} = m$ we should recover the instanton action for the $\mathbb{C}^2/\mathbb{Z}_2$ singularity which is known by perturbative means. It is very pleasing to see that this is indeed the case.
The bosonic part of the instanton action is easily handled. Just like we did
in going to the SPP, the further higgsing gives a mass to the difference
$\omega_{33} - \omega_{22}$ and $\bar\omega_{33} - \bar\omega_{22}$, allowing
us to set $\omega_{33} = \omega_{22}$ and $\bar\omega_{33} =
\bar\omega_{22}$. Since no new mass term for the chiral superfields is induced
in this process, we simply make this identification in the bosonic
action ({\ref{SboseSPP}) to obtain the well known orbifold result. Similarly,
  higgsing in the anti-holomorphic part of the action gives a mass to
  $\mu_{32}$, $\bar\mu_{32}$, and to the linear combinations $\mu_{33} -
  \mu_{22}$ and $\bar\mu_{33} - \bar\mu_{22}$ allowing us to set them to zero
  in both the anti-holomorphic and holomorphic fermionic actions
  (\ref{SantiSPP}) and (\ref{SholoSPP}). More interestingly, the fields
  $\Phi_{32}$ and $s_{32}$  appear only in the quartic part of the holomorphic
  action (\ref{SholoSPP}) and their VEV reduces these terms to the cubic ones expected in the orbifold case. Not only that, this last fact indicates that the quartic terms \emph{must} be present in the SPP case since without them we would not recover all the couplings for the $\mathbb{C}^2/\mathbb{Z}_2$ orbifold.

Now that we trust the action for SPP we can make another higgsing, this time
to the conifold theory. The conifold case is quite dramatic in that the only
allowed holomorphic terms in the fermionic action are quartic and if they were
not present there would be no hope of recovering the instanton action of $\mathcal{N}=4$ by further higgsing. On the other hand, by keeping these terms, one easily sees that further higgsing reduces to the desired action. To summarize this step, recall that we obtained (\ref{wConifold}) from (\ref{wSPP}) by higgsing $\Phi_{34} = -s_{34} = m$. Let us focus on the fermionic part of the action, since the bosonic part always works in the same way (here, $\omega_{11}=\omega_{33}$ and $\bar\omega_{11}=\bar\omega_{33}$). From the anti-holomorphic piece we have $\mu_{34}=0$ $\bar\mu_{34}=0$, $\mu_{11} = \mu_{33}$ and $\bar\mu_{11} = \bar\mu_{33}$, whereas from the holomorphic piece we can solve:
\beqs
        \mu _{13}= \frac{1}{m}\big(\Phi _{12}\mu _{24}- \mu_{12}s_{24}\big), &\quad&
       \bar{\mu }_{13} = \frac{1}{m}\big(\bar{\mu }_{12}\Phi _{24}-
s_{12}\bar{\mu}_{24}\big), \nonumber\\
           \mu _{41} = \frac{1}{m}\big(\Phi_{32}\mu _{23}-\mu _{32}s_{23}\big), &\quad&
     \bar{\mu}_{41}=  \frac{1}{m}\big(\bar{\mu }_{32}\Phi_{23}-s_{32}
\bar{\mu }_{23}\big).
\eeqs
Replacing these values in the fermionic action we obtain the complete instanton action for the conifold:

\beqs
S_{con}^{\omega} &=& \big( \bar \omega_{11} \Phi_{12} + s_{12} \bar \omega_{22}\big)
               \big(\Phi^\dagger_{21} \omega_{11} + \omega_{22} s^\dagger_{21}\big) +
               \big( \bar \omega_{22} \Phi_{21}^\dagger + s_{21}^\dagger \bar \omega_{11}\big)
               \big(\Phi_{12} \omega_{22} + \omega_{11} s_{12}\big)
               \nonumber \\
            &+&\big( \bar \omega_{11} \Phi_{32} + s_{32} \bar \omega_{22}\big)
               \big(\Phi^\dagger_{23} \omega_{11} + \omega_{22} s^\dagger_{23}\big) +
               \big( \bar \omega_{22} \Phi_{23}^\dagger + s_{23}^\dagger \bar \omega_{11}\big)
               \big(\Phi_{32} \omega_{22} + \omega_{11} s_{32}\big)
                \nonumber \\
            &+&\big( \bar \omega_{22} \Phi_{23} + s_{23} \bar \omega_{11}\big)
               \big(\Phi^\dagger_{32} \omega_{22} + \omega_{11} s^\dagger_{32}\big) +
               \big( \bar \omega_{11} \Phi_{32}^\dagger + s_{32}^\dagger \bar \omega_{22}\big)
               \big(\Phi_{23} \omega_{11} + \omega_{22} s_{23}\big)\nonumber \\
            &+&\big( \bar \omega_{22} \Phi_{24} + s_{24} \bar \omega_{11}\big)
               \big(\Phi^\dagger_{42} \omega_{22} + \omega_{11} s^\dagger_{42}\big) +
               \big( \bar \omega_{11} \Phi_{42}^\dagger + s_{42}^\dagger \bar \omega_{22}\big)
               \big(\Phi_{24} \omega_{11} + \omega_{22} s_{24}\big) \nn \\ &&
\eeqs

\beqs
S_{con}^{\overline{\mathrm{holo}}} &=&
 \big(\bar \mu_{12} \Phi^\dagger_{21} + \bar \mu_{32} \Phi^\dagger_{23}\big)\mu_{11} +
 \big(\bar \mu_{23} \Phi^\dagger_{32} + \bar \mu_{24} \Phi^\dagger_{42}\big)\mu_{22} +
 \big(s^\dagger_{32}\bar \mu_{23} + s^\dagger_{42}\bar \mu_{24} \big)\mu_{11} \nn \\ && +
 \big(s^\dagger_{23}\bar \mu_{32} + s^\dagger_{21}\bar \mu_{12} \big)\mu_{22}
-\bar\mu_{11} \big( \Phi^\dagger_{32} \mu_{23} + \Phi^\dagger_{42} \mu_{24} \big)
-\bar\mu_{22} \big( \Phi^\dagger_{21} \mu_{12} + \Phi^\dagger_{23} \mu_{32} \big) \nn \\ &&
-\bar\mu_{11} \big( \mu_{12} s^\dagger_{21} +  \mu_{32} s^\dagger_{23}\big)
-\bar\mu_{22} \big( \mu_{23} s^\dagger_{32} +  \mu_{24} s^\dagger_{42}\big)\label{SantiConi}
\eeqs

\beqs
S_{con}^{\mathrm{holo}} &=& \frac{1}{m} \big( \bar{\mu }_{12} \mu_{23} s_{32} s_{24}-\bar{\mu }_{12} \Phi_{23} \mu_{32} s_{24}+\bar{\mu }_{12} \Phi_{23} \Phi _{32} \mu _{24}+\bar{\mu}_{12} \Phi _{24} \mu_{32} s_{23} \nn \\&& -\bar{\mu}_{12} \Phi _{24} \Phi _{32} \mu_{23}-\bar{\mu }_{23} \mu _{12} s_{24} s_{32} +\bar{\mu }_{23} \mu_{32} s_{24} s_{12}+\bar{\mu }_{23} \Phi _{12} \mu _{24} s_{32} \nn \\ && -\bar{\mu }_{23} \Phi_{12} \Phi _{24} \mu_{32}+\bar{\mu}_{23} \Phi _{32} \Phi _{24} \mu_{12}-\bar{\mu }_{23} \Phi _{32} \mu_{24} s_{12}+\bar{\mu}_{24} \mu_{12} s_{23} s_{32}\nn \\ &&-\bar{\mu }_{24} \mu_{32} s_{23} s_{12}-\bar{\mu }_{24} \Phi_{12} \mu _{23} s_{32}+\bar{\mu }_{24} \Phi_{12} \Phi _{23} \mu _{32}+\bar{\mu}_{24} \Phi _{32} \mu _{23} s_{12}\nn \\ &&-\bar{\mu}_{24} \Phi _{32} \Phi _{23} \mu_{12}-\bar{\mu }_{32} \mu_{23} s_{12} s_{24}+\bar{\mu }_{32} \mu_{24} s_{12} s_{23}+\bar{\mu }_{32} \Phi_{23} \mu _{12} s_{24}\nn \\ &&-\bar{\mu }_{32} \Phi_{23} \Phi _{12} \mu_{24}-\bar{\mu}_{32} \Phi _{24} \mu _{12} s_{23}+\bar{\mu}_{32} \Phi _{24} \Phi _{12}\mu _{23}-\bar{\mu}_{12}\mu_{24} s_{32} s_{23}  \big)\label{SholoConi}
\eeqs

To test this result, one can make the further higgsing to the $\mathcal{N}=4$ theory by letting
$\Phi_{24} = - s_{24} = m$. Again, no new chiral superfield or neutral mode need to be integrated out and simply setting to zero the charged zero modes that acquire a mass: $\omega_{44} - \omega_{22}$,  $\bar\omega_{44} - \bar\omega_{22}$, $\mu_{44} - \mu_{22}$,  $\bar\mu_{44} - \bar\mu_{22}$, $\mu_{24}$, $\bar\mu_{24}$ yields the $\mathcal{N}=4$ expression reviewed in section~3. Notice that no masses are generated in the holomorphic sector (\ref{SholoConi}) but this sector is crucial for the recovery of the last term in (\ref{s2}).

\section{Recovering the ADS superpotential from the non-holomorphic couplings}
\label{RecoverADS}
We now make a short digression to show that in order to recover the ADS superpotential in the simplest
cases, the couplings of the instanton moduli to the non-holomorphic
matter fields\footnote{Recall that the above distinction is meaningful since
the instanton essentially chooses one chirality over the other.}
must be of the form found above. Hence we are building up confidence in the
procedure used to obtain the actions in the previous section, and will
be able to apply it systematically to other quivers. This is actually
a first indication that the rules proposed in the beginning of this paper
are consistent with the effects that have to be described by
instantons in quiver gauge theories.

According to the rules inferred from the previous (and subsequent) examples,
let us compute the spectrum of zero modes and their couplings
when there is one fractional instanton sitting on a node corresponding
to a gauge group, i.e. the instanton wraps a cycle which is also wrapped
by some space-filling branes. The number of space-filling branes
determines the rank of the gauge group at the associated node.
Note that as far as space-filling branes are concerned, in order to cancel gauge anomalies, other nodes might necessarily
need to be turned on (this is of course true only for chiral quivers).
This is not true for the instantonic branes since, roughly, the tadpole
charge can escape through the space-time directions. Hence, there are
no restrictions in considering a single instanton on a node, even if
the quiver is chiral (we will consider this generic case in this
section).\footnote{We thank Matteo Bertolini and Angel Uranga for discussions
on these issues.}

We are considering a single instanton, hence in the neutral zero mode
sector we will have no $s$ moduli. If we denote by $a$ the index of the
node with an instanton, we will have only
$a^\mu_{aa}$, $M_{aa}$ and $\lambda_{aa}$ zero modes in this sector.
Since there is both a gauge group and an instanton at node $a$, we will
have a pair of charged massless bosonic modes $\omega_{aa}$ and $\bar\omega_{aa}$,
as well as the charged fermionic zero modes  $\mu_{aa}$ and $\bar\mu_{aa}$.
In addition, to each matter field connecting to node $a$, both incoming
$\Phi_{ba}$ and outgoing $\Phi_{ac}$ (with $b$ and $c$ running on the
nodes connected to node $a$ by incoming and outgoing arrows respectively),
there will be charged fermionic zero modes $\mu_{ba}$
and  $\bar \mu_{ac}$.

The couplings in which these modes will necessarily be involved are the
following:
\beq
S_\omega = \bar \omega_{aa} \big(\sum_c \Phi_{ac}\Phi_{ca}^\dagger +
\sum_b \Phi^\dagger_{ab} \Phi_{ba} \big) \omega_{aa},
\eeq
\beq
S_\mu^{\overline{\mathrm{holo}}}=
\sum_b \bar\mu_{aa} \Phi^\dagger_{ab}\mu_{ba} -
\sum_c \bar \mu_{ac} \Phi_{ca}^\dagger \mu_{aa}.
\eeq
Additionally, they can also be involved in holomorphic couplings,
if there are corresponding superpotential terms:
\beq
S_\mu^{\mathrm{holo}}=
\bar\mu_{ac} \Phi_{cd} \dots \Phi_{eb}\mu_{ba},
\eeq
where the number of matter fields in the couplings above is given
by the order of the corresponding superpotential term minus two.

We now integrate over all of the zero modes. The integral over the neutral
zero modes $a^\mu_{aa}$ and $M_{aa}$ just gives the integral over chiral
superspace that tells us that we are computing a superpotential term.
The integral over the $\lambda_{aa}$ zero modes brings down a fermionic
delta function implementing the constraint
\beq
\bar \mu_{aa} \omega_{aa} +\bar \omega_{aa} \mu_{aa} = 0.
\eeq
If we take the gauge group at node $a$ to be of rank $N$, then all the
zero modes in the equation above have actually $N$ components over which
we must sum (the
$\omega$ and $\bar \omega$ are additionally Lorentz spinors, so that there
are in total two fermionic constraints).

Let us now focus on the fermionic integration:
\beq
\int [{\cal D} \bar \mu_{aa}]^N[{\cal D}  \mu_{aa}]^N
[{\cal D}\bar\mu_{ac}]^{N'}
[{\cal D}\mu_{ba}]^{N'}(\bar \mu_{aa} \omega_{aa} +\bar \omega_{aa} \mu_{aa})^2
e^{-S_\mu^{\overline{\mathrm{holo}}}-S_\mu^{\mathrm{holo}}},
\eeq
where $N'$ is the sum of the ranks of the $U(N_b)$ and $U(N_c)$
gauge groups connected with node $a$. The two sums must coincide because
of anomaly cancellation, so that $N'$ is essentially the number of
flavors of the $U(N)$ gauge theory at node $a$.

Performing first the integration over the $\bar \mu_{aa}$ and $\mu_{aa}$
moduli, we see that a pair is soaked up by the fermionic constraint,
while the others are soaked up by pulling down $2(N-1)$ times
$S_\mu^{\overline{\mathrm{holo}}}$. Together with $2(N-1)$ powers
of anti-holomorphic matter fields, we also bring down $2(N-1)$
zero modes of the type  $\mu_{ba}$ and  $\bar \mu_{ac}$. It is clear
that if $N'<N-1$, the contribution will then vanish.

If on the other hand we concentrate on the case $N'=N-1$, where the standard
gauge theory analysis \cite{Affleck:1983mk} tells us that there should
be an effective superpotential generated by a one instanton contribution,
we immediately see that all the other zero modes $\mu_{ba}$ and
$\bar \mu_{ac}$ are also exactly soaked up in this process. This means
that the terms in $S_\mu^{\mathrm{holo}}$ are irrelevant to this
contribution.\footnote{When $N'\geq N$, the terms in $S_\mu^{\mathrm{holo}}$
might start playing a non-trivial role. This would be related to gauge
theory instantons in theories with additional singlet matter fields
coupling to the flavors. The study of such effects is beyond the scope
of the present paper.} Thus, the integration over fermionic moduli
leaves us with an expression with $2N'$ powers of the anti-holomorphic
flavors of the $SU(N)$ gauge group. The expression we get is the same
that appears in the ADHM construction \cite{Atiyah:1978ri} (see also
\cite{Dorey:2002ik,Bianchi:2007ft}) after integrating over
the fermionic moduli
\beq
\int d^4x d^2 \theta \det (\Phi_{ca}^\dagger \Phi_{ab}^\dagger).
\eeq

The integration over the bosonic charged zero modes is again  the
standard ADHM one, derived in a stringy context in
\cite{Akerblom:2006hx}. Hence, the anti-holomorphic
pieces in the numerator and in the denominator cancel, leaving
us with the familiar ADS contribution
\beq
\int d^4x d^2 \theta \frac{1}{\det (\Phi_{ba} \Phi_{ac})} \propto
\int d^4x d^2\theta  W_{ADS}.
\eeq
As it is clear from the above, this is completely general and applies
also to  chiral quivers such as the ones considered in the next section.
It is also a consistency check for the zero mode actions that we
computed before, and for the rules explained at the beginning of the
paper. In particular, it is crucial that the couplings of the charged
bosonic zero modes are all proportional to $\Phi^\dagger \Phi$, and the
couplings of the charged fermionic zero modes to the anti-holomorphic
sector are all linear in $\Phi^\dagger$. Had we kept the subleading
terms proportional to higher powers of $\Phi^\dagger$, we would be in
a situation where the  instantons on a node occupied by
a gauge theory would yield a contribution in disagreement with the
one computed through the gauge theory itself.

We thus take the results of this section as a further confirmation
that we are indeed taking the correct procedure to perform the
higgsing in the instanton sector.

\section{Higgsing $\mathbb{C}^3/\mathbb{Z}_3\times\mathbb{Z}_3$ to toric
del Pezzo's}
Finally we discuss the higgsing from an orbifold quiver to the
del Pezzo toric quivers known as $dP_1$, $dP_2$ and $dP_3$ which
have been often considered in recent research on quiver gauge theories,
since they contain basically all the features the latter
can display.

There are two reasons why we want to look at these cases in some details. The first is that models based on these  singularities have attracted some attention in the context of dynamical supersymmetry breaking and might even yield phenomenologically interesting models. The second is that all the models considered so far, obtained from the resolution of $\mathbb{C}^3/\mathbb{Z}_2\times\mathbb{Z}_2$ are non chiral and one might wonder if the procedure generalizes to the (more interesting) chiral theories. We will see that it does, and
hopefully this should convince the reader that the recipe we gave in the introduction on how to build the instanton action directly from the quiver data is quite general, so that one need not go through the
(rather painful) higgsing procedure for even larger quivers.

Schematically, the higgsing procedure we will follow is represented by the chain:
\beq
   \mathbb{C}^3/\mathbb{Z}_3\times\mathbb{Z}_3 \stackrel{3,12}{\to} dP_3 \stackrel{1,0}{\to} dP_2 \stackrel{1,0}{\to} dP_1 \stackrel{1,0}{\to} \mathbb{C}^3/\mathbb{Z}_3,
\eeq
where the numbers above the arrows represent the numbers of chiral superfields (and neutral bosonic modes) acquiring a VEV and the number of chiral superfields (and neutral bosonic modes) that need to be integrated out as a consequence of this~\footnote{One could further higgs the $\mathbb{C}^3/\mathbb{Z}_3$ to the conifold case providing yet a consistency check.}. We see that most of the work is concentrated in the first step which we now describe.

We begin by writing the expression for the superpotential of the $\mathbb{C}^3/\mathbb{Z}_3\times\mathbb{Z}_3$
theory:
\beqs
     W_{\mathbb{Z}_3\times\mathbb{Z}_3} &=& -\Phi _{13} \Phi _{34} \Phi _{41} +
     \Phi_{15} \Phi_{54} \Phi_{41} - \Phi_{15} \Phi_{52} \Phi_{21} -
     \Phi_{26} \Phi_{63} \Phi_{32} \nn \\ &&+ \Phi_{34} \Phi_{46} \Phi_{63} +
     \Phi_{26} \Phi_{65} \Phi_{52} + \Phi_{17} \Phi_{72} \Phi_{21} -
     \Phi_{46} \Phi_{67} \Phi_{74} \nn \\ &&+ \Phi_{28} \Phi_{83} \Phi_{32} -
     \Phi_{48} \Phi_{85} \Phi_{54} - \Phi_{28} \Phi_{87} \Phi_{72} +
     \Phi_{48} \Phi_{87} \Phi_{74} \nn \\ &&+ \Phi_{13} \Phi_{39} \Phi_{91} -
     \Phi_{17} \Phi_{79} \Phi_{91} - \Phi_{59} \Phi_{96} \Phi_{65} +
     \Phi_{67} \Phi_{79} \Phi_{96} \nn \\ &&- \Phi_{39} \Phi_{98} \Phi_{83} +
     \Phi_{59} \Phi_{98} \Phi_{85} . \label{wZ3Z3}
\eeqs

To go to the $dP_3$ model we need to remove three nodes from the quiver
 diagram, that is higgs three chiral superfields.
 We follow \cite{Beasley:2001zp} and higgs $\Phi_{83} = \Phi_{79} = \Phi_{41} = m$.\footnote{
As it is well known, the higher del Pezzo's possess more than one ``toric
 phase". We will limit ourselves here to considering arguably the most natural
 one for $dP_3$ denoted by model I in \cite{Beasley:2001zp}.}
Substituting into (\ref{wZ3Z3}) we see that this gives a mass to twelve fields that need to be integrated out:
\beqs
&&\Phi_{32}=\frac{1}{m}\Phi_{87}\Phi_{72}, \quad
\Phi_{28}=\frac{1}{m}\Phi_{26}\Phi_{63}, \quad
\Phi_{34}=\frac{1}{m}\Phi_{39}\Phi_{91}, \nn \\
&&\Phi_{13}=\frac{1}{m}\Phi_{46}\Phi_{63}, \quad
\Phi_{54}=\frac{1}{m}\Phi_{52}\Phi_{21}, \quad
\Phi_{15}=\frac{1}{m}\Phi_{48}\Phi_{85}, \nn \\
&&\Phi_{91}=\frac{1}{m}\Phi_{72}\Phi_{21}, \quad
\Phi_{17}=\frac{1}{m}\Phi_{13}\Phi_{39}, \quad
\Phi_{96}=\frac{1}{m}\Phi_{74}\Phi_{46}, \nn \\
&&\Phi_{67}=\frac{1}{m}\Phi_{65}\Phi_{59}, \quad
\Phi_{98}=\frac{1}{m}\Phi_{91}\Phi_{13},  \quad
\Phi_{39}=\frac{1}{m}\Phi_{85}\Phi_{59}~,\label{intoutZ3Z3}
\eeqs
see Figure \ref{dP3D}. 
\begin{figure}
\begin{center}
\vspace{4cm}
\includegraphics[width=65mm]{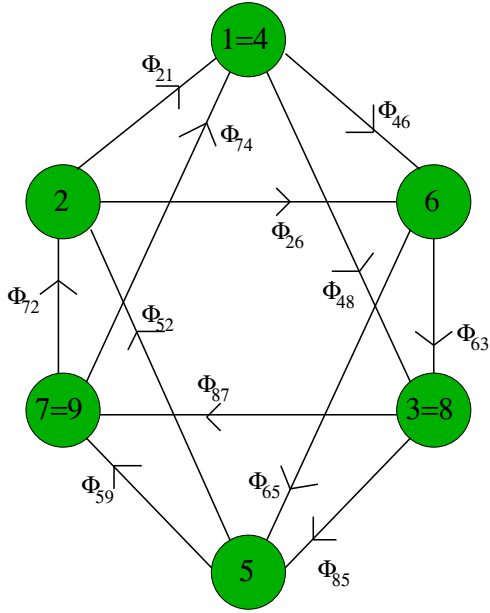}
\caption{\small{The $dP_3$ theory higgsed down from the $\mathbb{Z}_3\times\mathbb{Z}_3$ theory.}}
\label{dP3D}
\end{center}
\end{figure}
Note that some fields are expressed in terms
of other fieds which are themselves integrated out. This results in some
expressions being quartic in the left-over matter fields:
\beq
\Phi_{34}=\frac{1}{m^3} \Phi_{85}\Phi_{59}\Phi_{72}\Phi_{21}, \quad
\Phi_{17}=\frac{1}{m^3}\Phi_{46}\Phi_{63}\Phi_{85}\Phi_{59}, \quad
\Phi_{98}=\frac{1}{m^3}\Phi_{72}\Phi_{21}\Phi_{46}\Phi_{63}.
\label{quarticsub}
\eeq

The resulting superpotential for the $dP_3$ model is thus given by
\beqs
W_{dP_3} &=& \frac{1}{m^3} \big(\Phi _{21} \Phi _{46} \Phi _{63}\Phi _{85} \Phi _{59} \Phi _{72} \big) \nn \\ &&-
    \frac{1}{m} \big(\Phi _{46} \Phi _{65} \Phi _{59}\Phi _{74} +
         \Phi _{21} \Phi _{48} \Phi _{85}\Phi _{52} + \Phi _{26} \Phi _{63} \Phi _{87}\Phi _{72}\big) \nn \\ &&+
      \Phi _{26} \Phi _{65} \Phi _{52} +  \Phi _{48} \Phi _{87} \Phi _{74}. \label{wdP3}
\eeqs

The action for the charged bosonic zero modes is obtained, as before, by setting to zero the massive modes, i.e. identifying $\omega_{88} = \omega_{33}$, $\omega_{99} = \omega_{77}$, $\omega_{44} = \omega_{11}$, and dropping the $1/|m|^2$ terms that we claim must be suppressed in the IR limit in order to recover the right actions by further higgsing.
This results in the following rather unwieldy expression:

\beqs
S_{dP_3}^{\omega} &=& \big( \bar \omega_{22} \Phi_{26} + s_{26} \bar \omega_{66}\big)
               \big(\Phi^\dagger_{62} \omega_{22} + \omega_{66} s^\dagger_{62}\big)
             + \big( \bar \omega_{66} \Phi^\dagger_{62} + s^\dagger_{62} \bar \omega_{22}\big)
               \big(\Phi_{26}\omega_{66} + \omega_{22} s_{26}\big) \nn \\ &&
               \big( \bar \omega_{55} \Phi_{52} + s_{52} \bar \omega_{22}\big)
               \big(\Phi^\dagger_{25} \omega_{55} + \omega_{22} s^\dagger_{25}\big)
             + \big( \bar \omega_{22} \Phi^\dagger_{25} + s^\dagger_{25} \bar \omega_{55}\big)
               \big(\Phi_{52}\omega_{22} + \omega_{55} s_{52}\big) \nn \\ &&
               \big( \bar \omega_{66} \Phi_{65} + s_{65} \bar \omega_{55}\big)
               \big(\Phi^\dagger_{56} \omega_{66} + \omega_{55} s^\dagger_{56}\big)
             + \big( \bar \omega_{55} \Phi^\dagger_{56} + s^\dagger_{56} \bar \omega_{66}\big)
               \big(\Phi_{65}\omega_{55} + \omega_{66} s_{65}\big) \nn \\ &&
               \big( \bar \omega_{11} \Phi_{46} + s_{46} \bar \omega_{66}\big)
               \big(\Phi^\dagger_{64} \omega_{11} + \omega_{66} s^\dagger_{64}\big)
             + \big( \bar \omega_{66} \Phi^\dagger_{64} + s^\dagger_{64} \bar \omega_{11}\big)
               \big(\Phi_{46}\omega_{66} + \omega_{11} s_{46}\big) \nn \\ &&
               \big( \bar \omega_{55} \Phi_{59} + s_{59} \bar \omega_{77}\big)
               \big(\Phi^\dagger_{95} \omega_{55} + \omega_{77} s^\dagger_{95}\big)
             + \big( \bar \omega_{77} \Phi^\dagger_{95} + s^\dagger_{95} \bar \omega_{55}\big)
               \big(\Phi_{59}\omega_{77} + \omega_{55} s_{59}\big) \nn \\ &&
               \big( \bar \omega_{77} \Phi_{74} + s_{74} \bar \omega_{11}\big)
               \big(\Phi^\dagger_{47} \omega_{77} + \omega_{11} s^\dagger_{47}\big)
             + \big( \bar \omega_{11} \Phi^\dagger_{47} + s^\dagger_{47} \bar \omega_{77}\big)
               \big(\Phi_{74}\omega_{11} + \omega_{77} s_{74}\big) \nn \\ &&
               \big( \bar \omega_{22} \Phi_{21} + s_{21} \bar \omega_{11}\big)
               \big(\Phi^\dagger_{12} \omega_{22} + \omega_{11} s^\dagger_{12}\big)
             + \big( \bar \omega_{11} \Phi^\dagger_{12} + s^\dagger_{12} \bar \omega_{22}\big)
               \big(\Phi_{21}\omega_{11} + \omega_{22} s_{21}\big) \nn \\ &&
               \big( \bar \omega_{11} \Phi_{48} + s_{48} \bar \omega_{33}\big)
               \big(\Phi^\dagger_{84} \omega_{11} + \omega_{33} s^\dagger_{84}\big)
             + \big( \bar \omega_{33} \Phi^\dagger_{84} + s^\dagger_{84} \bar \omega_{11}\big)
               \big(\Phi_{48}\omega_{33} + \omega_{11} s_{48}\big) \nn \\ &&
               \big( \bar \omega_{33} \Phi_{85} + s_{85} \bar \omega_{55}\big)
               \big(\Phi^\dagger_{58} \omega_{33} + \omega_{55} s^\dagger_{58}\big)
             + \big( \bar \omega_{55} \Phi^\dagger_{58} + s^\dagger_{58} \bar \omega_{33}\big)
               \big(\Phi_{85}\omega_{55} + \omega_{33} s_{85}\big) \nn \\ &&
               \big( \bar \omega_{66} \Phi_{63} + s_{63} \bar \omega_{33}\big)
               \big(\Phi^\dagger_{36} \omega_{66} + \omega_{33} s^\dagger_{36}\big)
             + \big( \bar \omega_{33} \Phi^\dagger_{36} + s^\dagger_{36} \bar \omega_{66}\big)
               \big(\Phi_{63}\omega_{33} + \omega_{66} s_{63}\big) \nn \\ &&
               \big( \bar \omega_{77} \Phi_{72} + s_{72} \bar \omega_{22}\big)
               \big(\Phi^\dagger_{27} \omega_{77} + \omega_{22} s^\dagger_{27}\big)
             + \big( \bar \omega_{22} \Phi^\dagger_{27} + s^\dagger_{27} \bar \omega_{77}\big)
               \big(\Phi_{72}\omega_{22} + \omega_{77} s_{72}\big) \nn \\ &&
               \big( \bar \omega_{33} \Phi_{87} + s_{87} \bar \omega_{77}\big)
               \big(\Phi^\dagger_{78} \omega_{33} + \omega_{77} s^\dagger_{78}\big)
             + \big( \bar \omega_{77} \Phi^\dagger_{78} + s^\dagger_{78} \bar \omega_{33}\big)
               \big(\Phi_{87}\omega_{77} + \omega_{33} s_{87}\big).\nn \\ && \label{sbosedP3}
\eeqs

Expression (\ref{sbosedP3}) can be understood by noticing that to every chiral field surviving the higgsing procedure there corresponds a line in the expression, coupling this fields (and the corresponding neutral mode) to the two charged bosonic zero modes emanating from the nodes connected by the chiral field:
\beq
     \big( \bar \omega_{bb} \Phi_{ba} + s_{ba} \bar \omega_{aa}\big)\big(\Phi^\dagger_{ab}\omega_{bb} + \omega_{aa} s^\dagger_{ab}\big)
    +\big( \bar \omega_{aa} \Phi^\dagger_{ab} + s^\dagger_{ab} \bar \omega_{bb}\big)\big(\Phi_{ba}\omega_{aa} + \omega_{bb} s_{ba}\big)
     \label{sbosenutshell}
\eeq
Expression (\ref{sbosedP3}) is nothing but a repetition of (\ref{sbosenutshell}) for each chiral field, where in some places we have also replaced the bosonic modes that have been integrated out to avoid redundancy, (e.g. in the fourth line of (\ref{sbosedP3}) we write $\omega_{11}$ instead of $\omega_{44}$).

In the fermionic action, the following zero modes are made massive by the higgsing procedure.
From the anti-holomorphic piece:
\beqs
   && \mu_{88}-\mu_{33}, ~~ \mu_{99}-\mu_{77}, ~~ \mu_{44}-\mu_{11}, ~~ \mu_{83}, ~~ \mu_{79},  ~~ \mu_{41}, \nn \\
   && \bar\mu_{88}-\bar\mu_{33}, ~~ \bar\mu_{99}-\bar\mu_{77}, ~~ \bar\mu_{44}-\bar\mu_{11}, ~~
       \bar\mu_{83}, ~~ \bar\mu_{79},  ~~ \bar\mu_{41}, \label{fermidP3anti}
\eeqs
and from the holomorphic piece:
\beqs
   && \mu_{13}, ~~ \mu_{34}, ~~ \mu_{15}, ~~ \mu_{54}, ~~ \mu_{17}, ~~ \mu_{91}, ~~ \mu_{28}, ~~
      \mu_{32}, ~~ \mu_{39}, ~~ \mu_{98}, ~~ \mu_{67}, ~~ \mu_{96}, \nn \\
   && \bar\mu_{13}, ~~ \bar\mu_{34}, ~~ \bar\mu_{15}, ~~ \bar\mu_{54}, ~~ \bar\mu_{17}, ~~ \bar\mu_{91},
     ~~ \bar\mu_{28}, ~~ \bar\mu_{32}, ~~ \bar\mu_{39}, ~~ \bar\mu_{98}, ~~ \bar\mu_{67}, ~~ \bar\mu_{96}. \label{fermidP3holo}
\eeqs
It is easy to see what fields become massive in the holomorphic case by looking at the superpotential (\ref{wZ3Z3}) and recalling that the holomorphic actions for the fermionic zero modes has the same structure. Thus, the fermions that become massive are those with the same index structure of the fields in (\ref{intoutZ3Z3}). What is more interesting is that, integrating out the modes in the holomorphic and anti-holomorphic action separately (in order to preserve the holomorphic/anti-holomorphic decoupling) sets the fields in ({\ref{fermidP3anti}) to zero and solves the ones in ({\ref{fermidP3holo}) in terms of purely holomorphic quantities. Substituting back into the orbifold action and taking the scaling limit discussed previously we obtain the following complete action for the fermionic charged zero modes on the $dP_3$ theory:

\beqs
S_{dP_3}^{\mathrm{holo}} &=& \frac{1}{m^3} \big(  \bar\mu_{21} \mu_{46} s_{63} s_{85} s_{59} s_{72}  -
                                               \bar\mu_{21} \Phi_{46} \mu_{63} s_{85} s_{59} s_{72}  +
                                               \bar\mu_{21} \Phi_{46} \Phi_{63} \mu_{85} s_{59} s_{72} \nn \\ && -
                                               \bar\mu_{21} \Phi_{46} \Phi_{63} \Phi_{85} \mu_{59} s_{72}  +
                                               \bar\mu_{21} \Phi_{46} \Phi_{63} \Phi_{85} \Phi_{59} \mu_{72}  +
                                               \bar\mu_{46} \mu_{63} s_{85} s_{59} s_{72} s_{21} \nn \\ && -
                                               \bar\mu_{46} \Phi_{63} \mu_{85} s_{59} s_{72} s_{21}  +
                                               \bar\mu_{46} \Phi_{63} \Phi_{85} \mu_{59} s_{72} s_{21}  -
                                               \bar\mu_{46} \Phi_{63} \Phi_{85} \Phi_{59} \mu_{72} s_{21} \nn \\ && +
                                               \bar\mu_{46} \Phi_{63} \Phi_{85} \Phi_{59} \Phi_{72} \mu_{21}  +
                                               \bar\mu_{63} \mu_{85} s_{59} s_{72} s_{21} s_{46}  -
                                               \bar\mu_{63} \Phi_{85} \mu_{59} s_{72} s_{21} s_{46}\nn \\ &&  +
                                               \bar\mu_{63} \Phi_{85} \Phi_{59} \mu_{72} s_{21} s_{46}  -
                                               \bar\mu_{63} \Phi_{85} \Phi_{59} \Phi_{72} \mu_{21} s_{46}  +
                                               \bar\mu_{63} \Phi_{85} \Phi_{59} \Phi_{72} \Phi_{21} \mu_{46} \nn \\ && +
                                               \bar\mu_{85} \mu_{59} s_{72} s_{21} s_{46} s_{63}  -
                                               \bar\mu_{85} \Phi_{59} \mu_{72} s_{21} s_{46} s_{63}  +
                                               \bar\mu_{85} \Phi_{59} \Phi_{72} \mu_{21} s_{46} s_{63} \nn \\ && -
                                               \bar\mu_{85} \Phi_{59} \Phi_{72} \Phi_{21} \mu_{46} s_{63}  +
                                               \bar\mu_{85} \Phi_{59} \Phi_{72} \Phi_{21} \Phi_{46} \mu_{63}  +
                                               \bar\mu_{59} \mu_{72} s_{21} s_{46} s_{63} s_{85} \nn \\ && -
                                               \bar\mu_{59} \Phi_{72} \mu_{21} s_{46} s_{63} s_{85}  +
                                               \bar\mu_{59} \Phi_{72} \Phi_{21} \mu_{46} s_{63} s_{85}  -
                                               \bar\mu_{59} \Phi_{72} \Phi_{21} \Phi_{46} \mu_{63} s_{85} \nn \\ && +
                                               \bar\mu_{59} \Phi_{72} \Phi_{21} \Phi_{46} \Phi_{63} \mu_{85}  +
                                               \bar\mu_{72} \mu_{21} s_{46} s_{63} s_{85} s_{59}  -
                                               \bar\mu_{72} \Phi_{21} \mu_{46} s_{63} s_{85} s_{59} \nn \\ && +
                                               \bar\mu_{72} \Phi_{21} \Phi_{46} \mu_{63} s_{85} s_{59}  -
                                               \bar\mu_{72} \Phi_{21} \Phi_{46} \Phi_{63} \mu_{85} s_{59}  +
                                               \bar\mu_{72} \Phi_{21} \Phi_{46} \Phi_{63} \Phi_{85} \mu_{59} \big) \nn \\ &&
                                              - \frac{1}{m} \big( \bar\mu_{46} \mu_{65} s_{59} s_{74} -
                                                                 \bar\mu_{46} \Phi_{65} \mu_{59} s_{74} +
                                                                 \bar\mu_{46} \Phi_{65} \Phi_{59} \mu_{74} +
                                                                 \bar\mu_{65} \mu_{59} s_{74} s_{46} \nn \\ &&-
                                                                 \bar\mu_{65} \Phi_{59} \mu_{74} s_{46} +
                                                                 \bar\mu_{65} \Phi_{59} \Phi_{74} \mu_{46} +
                                                                 \bar\mu_{59} \mu_{74} s_{46} s_{65} -
                                                                 \bar\mu_{59} \Phi_{74} \mu_{46} s_{65}\nn \\ && +
                                                                 \bar\mu_{59} \Phi_{74} \Phi_{46} \mu_{65} +
                                                                 \bar\mu_{74} \mu_{46} s_{65} s_{59} -
                                                                 \bar\mu_{74} \Phi_{46} \mu_{65} s_{59} +
                                                                 \bar\mu_{74} \Phi_{46} \Phi_{65} \mu_{59} \nn \\ &&+
                                                                 \bar\mu_{21} \mu_{48} s_{85} s_{52} -
                                                                 \bar\mu_{21} \Phi_{48} \mu_{85} s_{52} +
                                                                 \bar\mu_{21} \Phi_{48} \Phi_{85} \mu_{52} +
                                                                 \bar\mu_{48} \mu_{85} s_{52} s_{21}\nn \\ && -
                                                                 \bar\mu_{48} \Phi_{85} \mu_{52} s_{21} +
                                                                 \bar\mu_{48} \Phi_{85} \Phi_{52} \mu_{21} +
                                                                 \bar\mu_{85} \mu_{52} s_{21} s_{48} -
                                                                 \bar\mu_{85} \Phi_{52} \mu_{21} s_{48} \nn \\ &&+
                                                                 \bar\mu_{85} \Phi_{52} \Phi_{21} \mu_{48} +
                                                                 \bar\mu_{52} \mu_{21} s_{48} s_{85} -
                                                                 \bar\mu_{52} \Phi_{21} \mu_{48} s_{85} +
                                                                 \bar\mu_{52} \Phi_{21} \Phi_{48} \mu_{85}\nn \\ && +
                                                                 \bar\mu_{26} \mu_{63} s_{87} s_{72} -
                                                                 \bar\mu_{26} \Phi_{63} \mu_{87} s_{72} +
                                                                 \bar\mu_{26} \Phi_{63} \Phi_{87} \mu_{72} +
                                                                 \bar\mu_{63} \mu_{87} s_{72} s_{26} \nn \\ &&-
                                                                 \bar\mu_{63} \Phi_{87} \mu_{72} s_{26} +
                                                                 \bar\mu_{63} \Phi_{87} \Phi_{72} \mu_{26} +
                                                                 \bar\mu_{87} \mu_{72} s_{26} s_{63} -
                                                                 \bar\mu_{87} \Phi_{72} \mu_{26} s_{63} \nn \\ &&+
                                                                 \bar\mu_{87} \Phi_{72} \Phi_{26} \mu_{63} +
                                                                 \bar\mu_{72} \mu_{26} s_{63} s_{87} -
                                                                 \bar\mu_{72} \Phi_{26} \mu_{63} s_{87} +
                                                                 \bar\mu_{72} \Phi_{26} \Phi_{63} \mu_{87} \big)\nn \\ &&
                                  -\bar\mu_{26} \mu_{65} s_{52} + \bar\mu_{26} \Phi_{65} \mu_{52}
                                         -\bar\mu_{65} \mu_{52} s_{26} + \bar\mu_{65} \Phi_{52} \mu_{26}
                                         -\bar\mu_{52} \mu_{26} s_{65} + \bar\mu_{52} \Phi_{26} \mu_{65} \nn \\ &&
                                         -\bar\mu_{48} \mu_{87} s_{74} + \bar\mu_{48} \Phi_{87} \mu_{74}
                                         -\bar\mu_{87} \mu_{74} s_{48} + \bar\mu_{87} \Phi_{74} \mu_{48}
                                         -\bar\mu_{74} \mu_{48} s_{87} + \bar\mu_{74} \Phi_{48} \mu_{87} \nn \\ &&
\eeqs

\beqs
S_{dP_3}^{\overline{\mathrm{holo}}} &=&
 \big(\bar \mu_{46} \Phi^\dagger_{64} + \bar \mu_{48} \Phi^\dagger_{84}\big)\mu_{11} +
 \big(\bar \mu_{21} \Phi^\dagger_{12} + \bar \mu_{26} \Phi^\dagger_{62}\big)\mu_{22} +
 \big(\bar \mu_{85} \Phi^\dagger_{58} + \bar \mu_{87} \Phi^\dagger_{78}\big)\mu_{33} \nn \\&&+
 \big(\bar \mu_{52} \Phi^\dagger_{25} + \bar \mu_{59} \Phi^\dagger_{95}\big)\mu_{55} +
 \big(\bar \mu_{63} \Phi^\dagger_{36} + \bar \mu_{65} \Phi^\dagger_{56}\big)\mu_{66} +
 \big(\bar \mu_{72} \Phi^\dagger_{27} + \bar \mu_{74} \Phi^\dagger_{47}\big)\mu_{77} \nn \\&&+
 \big(s^\dagger_{12}\bar \mu_{21} + s^\dagger_{47}\bar \mu_{74} \big)\mu_{11} +
 \big(s^\dagger_{25}\bar \mu_{52} + s^\dagger_{27}\bar \mu_{72} \big)\mu_{22} +
 \big(s^\dagger_{84}\bar \mu_{48} + s^\dagger_{36}\bar \mu_{63} \big)\mu_{33} \nn \\&&+
 \big(s^\dagger_{56}\bar \mu_{65} + s^\dagger_{58}\bar \mu_{85} \big)\mu_{55} +
 \big(s^\dagger_{62}\bar \mu_{26} + s^\dagger_{64}\bar \mu_{46} \big)\mu_{66} +
 \big(s^\dagger_{95}\bar \mu_{59} + s^\dagger_{78}\bar \mu_{87} \big)\mu_{77} \nn \\&&
-\bar\mu_{11} \big( \Phi^\dagger_{12} \mu_{21} + \Phi^\dagger_{47} \mu_{74} \big)
-\bar\mu_{22} \big( \Phi^\dagger_{25} \mu_{52} + \Phi^\dagger_{27} \mu_{72} \big)
-\bar\mu_{33} \big( \Phi^\dagger_{36} \mu_{63} + \Phi^\dagger_{84} \mu_{48} \big)\nn \\&&
-\bar\mu_{55} \big( \Phi^\dagger_{56} \mu_{65} + \Phi^\dagger_{58} \mu_{85} \big)
-\bar\mu_{66} \big( \Phi^\dagger_{62} \mu_{26} + \Phi^\dagger_{64} \mu_{46} \big)
-\bar\mu_{77} \big( \Phi^\dagger_{78} \mu_{87} + \Phi^\dagger_{95} \mu_{59} \big)\nn \\&&
-\bar\mu_{11} \big( \mu_{46} s^\dagger_{64} +  \mu_{48} s^\dagger_{84}\big)
-\bar\mu_{22} \big( \mu_{21} s^\dagger_{12} +  \mu_{26} s^\dagger_{62}\big)
-\bar\mu_{33} \big( \mu_{85} s^\dagger_{58} +  \mu_{87} s^\dagger_{78}\big)\nn \\&&
-\bar\mu_{55} \big( \mu_{52} s^\dagger_{25} +  \mu_{59} s^\dagger_{95}\big)
-\bar\mu_{66} \big( \mu_{63} s^\dagger_{36} +  \mu_{65} s^\dagger_{56}\big)
-\bar\mu_{77} \big( \mu_{72} s^\dagger_{27} +  \mu_{74} s^\dagger_{47}\big) \label{SantidP3}
\eeqs
Note that in order to recover all of the 30 sixth order terms above, it is
crucial that some fields have to be substituted by their quartic expressions
as in (\ref{quarticsub}), and similar expressions for the $s$ moduli.

We decided to present the full expression (which is not very practical in itself) because from now on the reader can easily convince herself that further higgsing yields the expressions for the lower del Pezzo's. In particular, setting $\Phi_{85} = - s_{85} = m$ yields one of the phases of the $dP_2$ theory, whose superpotential is simply obtained by substitution in (\ref{wdP3}) without the need of integrating anything out:
\beqs
W_{dP_2} &=& \frac{1}{m^2} \big(\Phi _{21} \Phi _{46} \Phi _{63}\Phi _{59} \Phi _{72} \big) -
    \frac{1}{m} \big(\Phi _{46} \Phi _{65} \Phi _{59}\Phi _{74} + \Phi _{26} \Phi _{63} \Phi _{87}\Phi _{72}\big) \nn \\ &&+
      \Phi _{26} \Phi _{65} \Phi _{52} +  \Phi _{48} \Phi _{87} \Phi _{74} -
         \Phi _{21} \Phi _{48}\Phi _{52}~,  \label{wdP2}
\eeqs
see Figure \ref{dP2D}.
\begin{figure}
\begin{center}
\includegraphics[width=75mm]{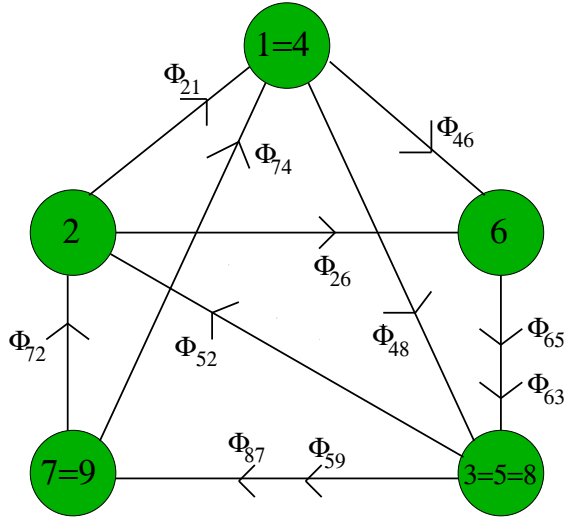}
\caption{\small{The $dP_2$ theory higgsed down from the $dP_3$ theory.}}
\label{dP2D}
\end{center}
\end{figure}

Similarly, substituting the VEVs in the bosonic and fermionic actions one can easily see that the fields $\omega_{88} - \omega_{55}$ and
$\bar\omega_{88} - \bar\omega_{55}$ become massive, allowing to eliminate one of the two elements in favor of the other and (from the anti-holomorphic piece) the fields
\beq
   \mu_{85}, ~~ \bar\mu_{85}, ~~ \mu_{88} - \mu_{55}, ~~ \bar\mu_{88} - \bar\mu_{55}
\eeq
become massive and are to be integrated out (set to zero). No further massive field arises from the holomorphic action and this is of course related to the fact that the structure of the holomorphic action reflects that of the superpotential.

We spare the reader the explicit expression for the instanton action in the $dP_2$ case that can be trivially retrieved from the comments above and move to the $dP_1$ model by further higgsing $\Phi_{72} = - s_{72} = m$. Once again, no chiral field acquires a mass and the superpotential for this (unique) toric phase is
\beqs
    W_{dP_1} &=& \frac{1}{m} \big(\Phi _{21} \Phi _{46} \Phi _{63}\Phi _{59} - \Phi _{46} \Phi _{65} \Phi _{59}\Phi _{74}\big)
                  \nn \\ &&
         - \Phi _{26} \Phi _{63} \Phi _{87} + \Phi _{26} \Phi _{65} \Phi _{52} +  \Phi _{48} \Phi _{87} \Phi _{74} -
         \Phi _{21} \Phi _{48}\Phi _{52}~, \label{wdP1}
\eeqs
see Figure \ref{dP1D}.
\begin{figure}
\begin{center}
\includegraphics[width=60mm]{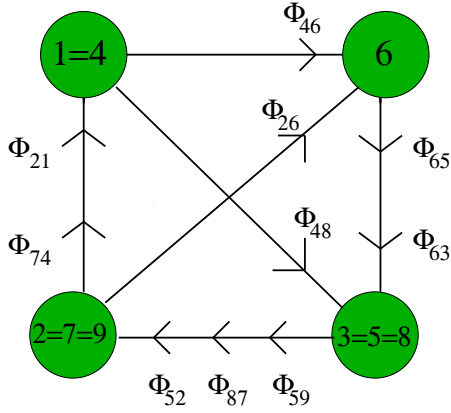}
\caption{\small{The $dP_1$ theory higgsed down from the $dP_2$ theory.}}
\label{dP1D}
\end{center}
\end{figure}

Just as in the previous step, the instanton action is trivially retrieved by making the appropriate substitutions:
\beq
   \mu_{72}=0, ~~ \bar\mu_{72}=0, ~~ \mu_{77} = \mu_{22}, ~~ \bar\mu_{77} = \bar\mu_{22}, ~~ \omega_{77} = \omega_{22},
   ~~ \bar\omega_{77} = \bar\omega_{22}
\eeq
enforcing the integrating out of the massive modes. We simply report the holomorphic part of the fermionic action for convenience of the reader and because it does have various applications.
\beqs
         S_{dP_1}^{\mathrm{holo}} &=& \frac{1}{m} \big(
              \bar\mu_{21}\mu_{46} s_{63} s_{59} - \bar\mu_{21}\Phi_{46} \mu_{63} s_{59} + \bar\mu_{21}\Phi_{46} \Phi_{63} \mu_{59} +
              \bar\mu_{46}\mu_{63} s_{59} s_{21} \nn \\ &&
                                - \bar\mu_{46}\Phi_{63} \mu_{59} s_{21} + \bar\mu_{46}\Phi_{63} \Phi_{59} \mu_{21} +
              \bar\mu_{63}\mu_{59} s_{21} s_{46} - \bar\mu_{63}\Phi_{59} \mu_{21} s_{46} \nn \\ &&
              + \bar\mu_{63}\Phi_{59} \Phi_{21} \mu_{46} +
              \bar\mu_{59}\mu_{21} s_{46} s_{63} - \bar\mu_{59}\Phi_{21} \mu_{46} s_{63} + \bar\mu_{59}\Phi_{21} \Phi_{46} \mu_{63}
              \nn \\ && -
              \bar\mu_{46}\mu_{65} s_{59} s_{74} + \bar\mu_{46}\Phi_{65} \mu_{59} s_{74} - \bar\mu_{46}\Phi_{65} \Phi_{59} \mu_{74} -
              \bar\mu_{65}\mu_{59} s_{74} s_{46} \nn \\ &&
                       + \bar\mu_{65}\Phi_{59} \mu_{74} s_{46} - \bar\mu_{65}\Phi_{59} \Phi_{74} \mu_{46} -
              \bar\mu_{59}\mu_{74} s_{46} s_{65} + \bar\mu_{59}\Phi_{74} \mu_{46} s_{65}\nn \\ &&
              - \bar\mu_{59}\Phi_{74} \Phi_{46} \mu_{65} -
              \bar\mu_{74}\mu_{46} s_{65} s_{59} + \bar\mu_{74}\Phi_{46} \mu_{65} s_{59} - \bar\mu_{74}\Phi_{46} \Phi_{65} \mu_{59} \big)
             \nn \\ &&   + \bar\mu_{26}\mu_{63} s_{87} - \bar\mu_{26}\Phi_{63} \mu_{87}
               + \bar\mu_{63}\mu_{87} s_{26} - \bar\mu_{63}\Phi_{87} \mu_{26}\nn \\ &&
               + \bar\mu_{87}\mu_{26} s_{63} - \bar\mu_{87}\Phi_{26} \mu_{63}
               - \bar\mu_{26}\mu_{65} s_{52} + \bar\mu_{26}\Phi_{65} \mu_{52}\nn \\ &&
               - \bar\mu_{65}\mu_{52} s_{26} + \bar\mu_{65}\Phi_{52} \mu_{26}
               - \bar\mu_{52}\mu_{26} s_{65} + \bar\mu_{52}\Phi_{26} \mu_{65}\nn \\ &&
               - \bar\mu_{48}\mu_{87} s_{74} + \bar\mu_{48}\Phi_{87} \mu_{74}
               - \bar\mu_{87}\mu_{74} s_{48} + \bar\mu_{87}\Phi_{74} \mu_{48}\nn \\ &&
               - \bar\mu_{74}\mu_{48} s_{87} + \bar\mu_{74}\Phi_{48} \mu_{87}
               + \bar\mu_{21}\mu_{48} s_{52} - \bar\mu_{21}\Phi_{48} \mu_{52}\nn \\ &&
               + \bar\mu_{48}\mu_{52} s_{21} - \bar\mu_{48}\Phi_{52} \mu_{21}
               + \bar\mu_{52}\mu_{21} s_{48} - \bar\mu_{52}\Phi_{21} \mu_{48} \label{SholodP1}
\eeqs

If our chain of derivation is correct, by further higgsing $\Phi_{46} = - s_{46} = m$ we should recover the action for the $\mathbb{C}^3/\mathbb{Z}_3$ orbifold. That this is indeed the case can be quickly ascertained by noticing that the quartic terms in (\ref{SholodP1}) always contain a term with index structure $(46)$. When such index is carried by a $\Phi$ or a $s$, the higgsing reduces it to a cubic term proper to the orbifold whereas, when the index falls on a $\mu$ or a $\bar\mu$ these terms are set to zero since those moduli get a mass from the anti-holomorphic term.

\section{Some applications and further directions}

As an illustration, in this section we present some simple applications of our
general results. We will be very sketchy and will not analyze the
dynamical consequences of the contributions we find, since that would
go beyond the scope of the present work. We merely want to present how
easily new contributions can be found by using the moduli actions derived
in the previous sections.

At this point, having left the general derivation and not having any further need of connecting different theories by higgsing, it is better to reconsider our previous decision and clean up the notation by relabeling the fields.

Let us start by studying the SPP gauge theory, where we have an arbitrary
number of fractional branes at node 1, a single spacefilling D-brane at node
2, node 3 unoccupied, and we put one instanton on node 2, see figure \ref{SPPwU1InstD}. This is an instance
of a $U(1)$ stringy instanton as discussed in \cite{Petersson:2007sc}.
\begin{figure}
\begin{center}
\includegraphics[width=60mm]{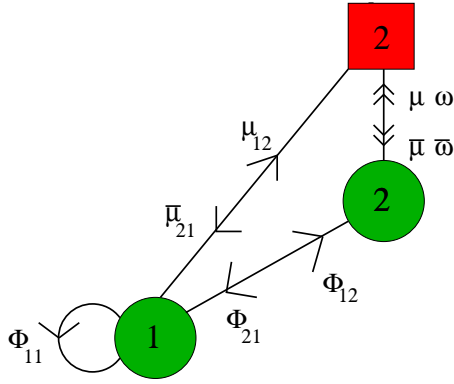}
\caption{\small{A U(1) instanton in the $SPP$ theory.}}
\label{SPPwU1InstD}
\end{center}
\end{figure}
There is one chiral superfield $\Phi_{11}$ in the adjoint representation of $U(N_1)$ and two bifundamental chiral superfields $\Phi_{12}$ and $\Phi_{21}$, transforming respectively in the $(\mathbf{ N}_1, \mathbf{ -})$ and $(\ov{\mathbf{ N}}_1, \mathbf{ +})$ of $U(N_1) \times U(1)$. The tree level superpotential for this configuration is given by the last term in (\ref{wSPP}), which reads, in the new notation:
\beqs
      W^{tree}_{SPP} =-\Phi_{21}\Phi_{11}\Phi_{12}~.
         \label{wSPPinst}
\eeqs

Let us begin with the  integral over the two neutral fermionic zero modes $\lambda_{\dot\alpha}$ which enforce the following fermionic ADHM constraints:
\beq
    \delta_F(\bar\omega_{\dot 1}\mu + \omega_{\dot 1}\bar\mu) \delta_F(\bar\omega_{\dot 2}\mu + \omega_{\dot 2}
    \bar\mu) = (\bar\omega_{\dot 1}\mu + \omega_{\dot 1}\bar\mu)(\bar\omega_{\dot 2}\mu + \omega_{\dot 2}
    \bar\mu) = \bar\omega_{\dot \alpha} \omega^{\dot \alpha}\mu \bar\mu
\eeq
since there are only two ``diagonal" fermionic zero-modes $\mu$ and $\bar\mu$.
This means that we cannot pull down any term in the anti-holomorphic action since they all include either a $\mu$ or a $\bar\mu$. The bosonic integral
\beq
     \int d^2\omega d^2\bar\omega \delta_B^3(\bar\omega \tau^c \omega)
     \bar\omega \omega \exp (-\bar\omega |\Phi|^2 \omega)
\eeq
turns out to be scale invariant \cite{Petersson:2007sc} (this is true only for the case of a $U(1)$
node) and thus gives a field independent non-zero multiplicative constant. (We have collectively denoted the chiral superfields by $\Phi$ in the exponent. The result is independent of $\Phi$ anyway.) Since the structure of the holomorphic couplings to the charged fermions is dictated by (\ref{wSPPinst}), we realize that there is only one term in the effective instanton action that remains to be integrated over:
\begin{equation}
 \int d^{N_1}\bar\mu_{21}d^{N_1} \mu_{12}~e^{\bar\mu_{21}\Phi_{11}\mu_{12}}~,
\end{equation}
and which yields a determinant of the field $\Phi_{11}$.
Thus, we have obtained the following contribution:
\begin{equation}
W^{inst}_{SPP}= \Lambda^{3-N_1}  \det [\Phi_{11}] ~,
\end{equation}
where we have lumped the numerical constants in the prefactor $\Lambda$, which
for our purposes can simply be viewed as a dimensionful parameter. It is
clear that one can hope to engineer in this way simple DSB models 
similar to the ones considered
in \cite{Aharony:2007db}.

Let us now turn to the $dP_1$ gauge theory at the bottom of the cascade, with
a fractional brane content given by $(N_1, N_2, N_3, N_4) =
(1,2,3,0)$. Treating the $SU(3)$ node as the gauge group we see that the
condition $N_f = N_c -1$ is satisfied and placing an instanton at this node
(see figure \ref{dP1wADSInstD}) one indeed generates the ADS superpotential, as discussed in section~6.
\begin{figure}
\begin{center}
\includegraphics[width=60mm]{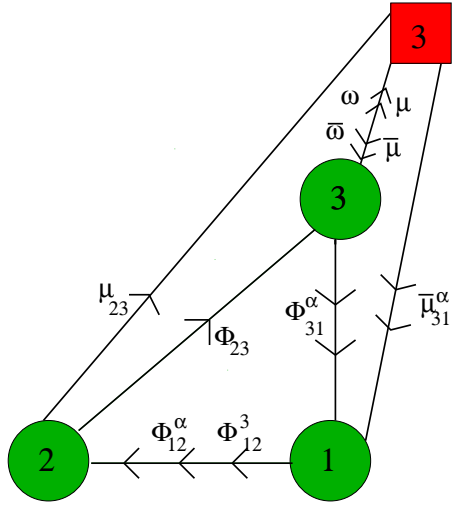}
\caption{\small{An ADS configuration in the $dP_1$ theory.}}
\label{dP1wADSInstD}
\end{center}
\end{figure}
Now one may want to consider more stringy phenomena such as what happens if one wraps an instanton at the unoccupied node or at the $U(1)$ node. It is easy to convince oneself that neither of  these configurations will give rise to any contribution. The instanton at the unoccupied node suffers from the usual problem with the presence of extra neutral fermionic zero modes that makes the whole expression vanish. The instanton at the $U(1)$ node instead has a \emph{charged} zero mode $\bar\mu_{12}^3$ not appearing anywhere in the action due to the fact that the tree level superpotential\footnote{The notation is such that $\alpha, \beta=1,2$ distinguish fields and moduli from the same nodes. In the case of fields from node 1 to node 2 we write $\Phi_{12}^\alpha$ and $\Phi_{12}^3$.}:
\beq
       W_{dP_1}^{tree} = \Phi_{23} \epsilon_{\alpha\beta} \Phi_{31}^\alpha \Phi_{12}^\beta
\eeq
does not contain the corresponding chiral field $\Phi_{12}^3$.
This makes its contribution vanish.

Let us instead see what happens when adding one regular brane to the picture, i.e. when the fractional brane content is $(N_1, N_2, N_3, N_4) = (2,3,4,1)$. This is the other case where we can have a $U(1)$ node with an instanton, see figure (\ref{dP1fullwU1InstD}).
\begin{figure}
\begin{center}
\includegraphics[width=73mm]{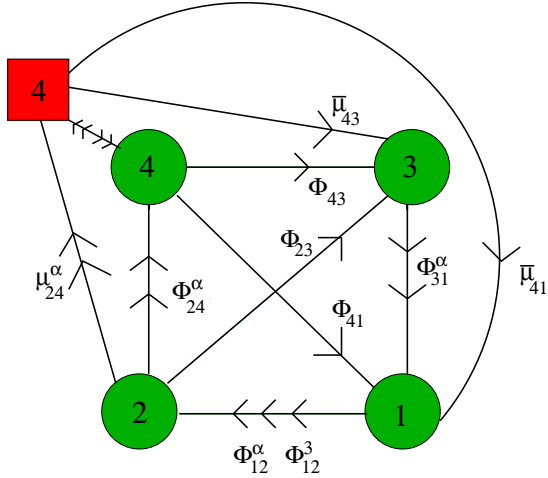}
\caption{\small{ A U(1) instanton configuration in the $dP_1$ theory.}}
\label{dP1fullwU1InstD}
\end{center}
\end{figure}
From (\ref{SholodP1}) we get the holomorphic couplings, (after relabeling)
\beq
    S_{dP_1}^{\mathrm{holo}} = \frac{1}{m}\big( \bar\mu_{43} \epsilon_{\alpha\beta}\Phi_{31}^\alpha \Phi_{12}^3\mu_{24}^\beta\big)
     -  \bar\mu_{41} \epsilon_{\alpha\beta}\Phi_{12}^\alpha\mu_{24}^\beta.
\eeq
The important difference in this configuration is that, since all chiral superfields appear in the tree level superpotential,
there will now be couplings in the instanton action that include all the fermionic moduli of this configuration.
Expanding the holomorphic action as to saturate the integral over all zero modes one can easily see that there is a contribution to the superpotential, albeit of high dimension.

As a final example, one can also consider a particular configuration in the $dP_3$ model. Here as well we relabel the fields in order to make the notation more intelligible. We have chosen the fractional brane assignment for the $dP_3$ theory to be $(N_1,N_2,N_3,N_4)=(P,M,P,M)$, see figure \ref{dP3U1InstD}, implying that we have removed the top and bottom nodes of figure \ref{dP3D}.
\begin{figure}
\begin{center}
\includegraphics[width=70mm]{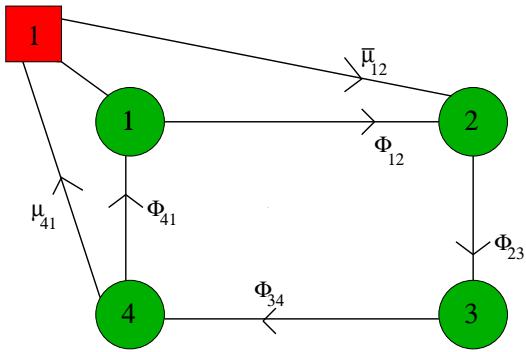}
\caption{\small{An U(1) instanton configuration for the $dP_3$ theory.}}
\label{dP3U1InstD}
\end{center}
\end{figure}
In the simplest possible case, we set $M=P=1$ and are left with a tree level superpotential given by:
\beqs
W_{dP_3}^{tree} =  \frac{1}{m}\Phi _{12} \Phi _{23} \Phi _{34}\Phi _{41},
\eeqs
where we note that all the chiral superfields in the quiver appear. This implies that if we place an instanton at the first node, there will be a coupling like $(1/m)\bar\mu _{12} \Phi _{23} \Phi _{34}\mu _{41}$ in the instanton action that will give rise to a quadratic superpotential term. Analogous mass terms will be generated if we instead place our instanton at a different node. Thus, summing over all possible locations for a single instanton we get the following structure,
\beq
       W_{dP_3}^{tot} = \frac{1}{m}\Phi _{12} \Phi _{23} \Phi _{34}\Phi _{41} +
       \frac{\Lambda_1^2}{m} \Phi _{23} \Phi _{34}+ \frac{\Lambda_2^2}{m} \Phi _{34} \Phi _{41} +
       \frac{\Lambda_3^2}{m}\Phi _{41} \Phi _{12} + \frac{\Lambda_4^2}{m} \Phi _{12} \Phi _{23} ~.
\eeq
Notice that we are not allowed to treat the ``$U(1)$" factors as gauge factors and this is also reflected in the fact that the mass terms generated are not invariant under this symmetries.
As a last remark about this case, notice that if we keep $P=1$ but go to $M>1$, from the two ``$U(1)$" instantons left we would get a vanishing contribution since the mass terms would be replaced by a determinant (e.g. $\mathrm{det}( \Phi_{12} \Phi_{23})$) of a matrix of rank one.

Clearly one can construct many more examples, particularly if one also allows for the presence of orientifolds. For example, the dynamical supersymmetry breaking configurations considered in \cite{Argurio:2006ny,Argurio:2007qk,Aharony:2007db}, which involved orbifolds/orientifolds of the conifold, can be obtained by the higgsing procedure since these singularities can all be embedded in an appropriate orbifold singularity. In summary, having at one's disposal the complete action for the instanton zero modes corresponding to any toric gauge theory should make this kind of investigation much more efficient and, hopefully, it will uncover corrections to the action of phenomenological relevance.

\section*{Acknowledgements}
It is a pleasure to thank Matteo Bertolini and Alberto Lerda for ongoing discussions and collaborations on related issues.
We thank Shamit Kachru for correspondence and sharing with us his results
prior to publication. Also, we would like to thank Jose F. Morales and Daniel Persson for stimulating discussions.
This work is partially supported by the European Commission FP6
Programme MRTN-CT-2004-005104, in which R.A. is associated to
V.U. Brussel, by IISN - Belgium
(convention 4.4505.86) and by the
``Interuniversity Attraction Poles Programme --Belgian Science Policy''.
R.A. is a Research Associate of the Fonds de la Recherche
Scientifique--F.N.R.S. (Belgium).
The research of G.F. is supported by the Swedish Research Council
(Vetenskapsr{\aa}det) contracts 622-2003-1124 and 621-2002-3884. Contract 622-2003-1124 also provides partial support for the research of C.P.

\end{document}